
\magnification=1200
\output={\plainoutput}

\newcount\pagenumber
\newcount\questionnumber
\newcount\sectionnumber
\newcount\appendixnumber
\newcount\equationnumber
\newcount\referencenumber

\def\ifundefined#1{\expandafter\ifx\csname#1\endcsname\relax}
\def\docref#1{\ifundefined{#1} {\bf ?.?}\message{#1 not yet defined,}
\else \csname#1\endcsname \fi}

\newread\bib
\newcount\linecount
\newcount\citecount
\newcount\localauthorcount
\def\article{\def\eqlabel##1{\edef##1{\sectionlabel.\the\equationnumber}}
\def\seclabel##1{\edef##1{\sectionlabel}}
\def\feqlabel##1{\ifnum\passcount=1
\immediate\write\crossrefsout{\relax}  
\immediate\write\crossrefsout{\def\string##1{\sectionlabel.
\the\equationnumber}}\else \fi }
\def\fseclabel##1{\ifnum\passcount=1
\immediate\write\crossrefsout{\relax}   
\immediate\write\crossrefsout{\def\string##1{\sectionlabel}}\else\fi}
\def\cite##1{\immediate\openin\bib=bib.tex\global\citecount=##1
\global\linecount=0{\loop\ifnum\linecount<\citecount \read\bib
to\temp \global\advance\linecount by 1\repeat\temp}\immediate\closein\bib}
\def\docite##1 auth ##2 title ##3 jour ##4 vol ##5 pages ##6 year ##7{
\par\noindent\item{\bf\the\referencenumber .}
 ##2, ##3, ##4, {\bf ##5}, ##6,
(##7).\par\vskip-0.8\baselineskip\noindent{
\global\advance\referencenumber by1}}
\def\dobkcite##1 auth ##2 title ##3 publisher ##4 year ##5{
\par\noindent\item{\bf\the\referencenumber .}
 ##2, {\it ##3}, ##4, (##5).
\par\vskip-0.8\baselineskip\noindent{\global\advance\referencenumber by1}}
\def\doconfcite##1 auth ##2 title ##3 conftitle ##4 editor ##5 publisher ##6
year ##7{
\par\noindent\item{\bf\the\referencenumber .}
##2, {\it ##3}, ##4,  {edited by: ##5}, ##6, (##7).
\par\vskip-0.8\baselineskip\noindent{\global\advance\referencenumber by1}}}

\def\appendixlabel{\ifcase\appendixnumber\or A\or B\or C\or D\or E\or
F\or G\or H\or I\or J\or K\or L\or M\or N\or O\or P\or Q\or R\or S\or
T\or U\or V\or W\or X\or Y\or Z\fi}

\def\sectionlabel{\ifnum\appendixnumber>0 \appendixlabel
\else\the\sectionnumber\fi}

\def\beginsection #1
 {{\global\subsecnumber=1\global\appendixnumber=0\global\advance\sectionnumber
by1}\equationnumber=1
\par\vskip 0.8\baselineskip plus 0.8\baselineskip
 minus 0.8\baselineskip
\noindent$\S$ {\bf \sectionlabel. #1}
\par\penalty 10000\vskip 0.6\baselineskip plus 0.8\baselineskip
minus 0.6\baselineskip \noindent}

\newcount\subsecnumber
\global\subsecnumber=1

\def\subsec #1 {\bf\par\vskip8truept  minus 8truept
\noindent \ifnum\appendixnumber=0 $\S\S\;$\else\fi
$\bf\sectionlabel.\the\subsecnumber$ #1
\global\advance\subsecnumber by1
\rm\par\penalty 10000\vskip6truept  minus 6truept\noindent}

\def\beginappendix #1
{{\global\subsecnumber=1\global\advance\appendixnumber
by1}\equationnumber=1\par
\vskip 0.8\baselineskip plus 0.8\baselineskip
 minus 0.8\baselineskip
\noindent
{\bf Appendix \appendixlabel . #1}
\par\vskip 0.8\baselineskip plus 0.8\baselineskip
 minus 0.8\baselineskip
\noindent}

\def\no{\eqno({\rm\sectionlabel}
.\the\equationnumber){\global\advance\equationnumber by1}}

\def\beginref #1 {\par\vskip 2.4 pt\noindent\item{\bf\the\referencenumber .}
\noindent #1\par\vskip 2.4 pt\noindent{\global\advance\referencenumber by1}}

\def\ref #1{{\bf [#1]}}

\def\specialbar{\bar}
\def\mod{\rm \,mod\,}
\def\dstar{d^*} \def\tr{\hbox{tr}\,}
\def\charav#1{\left\langle\chi^{#1}\right\rangle_p}

\def\tp{\tau_{+}'}
\def\tm{\tau_{-}'}
\def\Det{\hbox{Det}\;}

\article

\centerline{\bf Determinants of Laplacians, the Ray-Singer
Torsion on Lens Spaces}
\centerline{\bf and the Riemann zeta function}
\vskip1.25\baselineskip
\centerline{by}
\vskip1.25\baselineskip
\centerline{Charles Nash and D. J. O' Connor}
\par\vskip\baselineskip
\noindent
Department of Mathematical Physics\hfill School of Theoretical Physics
\par\noindent
St. Patrick's College\hfill              Dublin Institute for Advanced
Studies
\par\noindent
Maynooth\hfill                {\it and}\hfill  10 Burlington Road
\par\noindent
Ireland\hfill                            Dublin 4
\par\noindent
\null\hfill                              Ireland
\par\vskip3\baselineskip
\noindent
{{\bf Abstract}: We obtain explicit expressions for the determinants of the
Laplacians on zero and one forms for an infinite class of three dimensional
lens spaces $L(p,q)$. These expressions can be combined to obtain the
Ray-Singer torsion of these lens spaces. As a consequence we  obtain an
infinite class of formulae for the Riemann zeta function $\zeta(3)$. The
value of these determinants (and the torsion) grows as the size of the
fundamental group of the lens space increases and this is also computed.
The triviality of the torsion  for just the three lens spaces $L(6,1)$,
$L(10,3)$ and $L(12,5)$ is also noted.}
\beginsection{Introduction}
Topological phenomena are
  now known to play an important part in many
quantum field theories.  This is especially true of gauge theories.
There are also {\it topological quantum field theories} in which the
excitations are purely topological and the classical phase spaces of
these theories usually reduce to a finite dimensional space: these
spaces can be zero dimensional   discrete
sets, or whole moduli spaces. The semi-classical, or stationary phase,
approximation to the functional integral of such a theory is then
a weighted sum, or integral, over the {\it finite
dimensional} phase space. In addition, for some of these theories, this
approximation is {\it exact} providing thereby a reduction of the
functional integral to a finite dimensional integral.
\par
If a topological quantum field theory contains a gauge field
the reduced functional integral mentioned above
often consists of sums or integrals over {\it flat} connections; the
non-triviality of such connections is determined purely by their holonomy, and,
if $A$ is a flat connection over a manifold $M$, its holonomy is an
element of the fundamental group $\pi_1(M)$.
This means that an ideal laboratory within which to study
such theories is provided by taking the manifold $M$ to have a
non-trivial fundamental group but to be otherwise topologically rather
simple. An ideal way to do this is to take $M$ to be the quotient
of a sphere $S^n$ by a finite cyclic group $G$. This quotient $S^n/G$,
described in more detail below, is what is called a lens space, written as
$L(p,q)$.
In this paper we take $M$ to be a lens space on which is placed a topological
field theory whose classical phase space consists of flat connections.
\par
Our approach is to take the model
given by the field theory and analyse it in detail on a whole infinite
class of  lens spaces. We work in three dimensions and realise $M$
as the quotient of the  manifold $S^3$ by the action of a
discrete group $Z_{p}$. The resulting partition function on this
manifold is a combinatorial invariant of the manifold known as the
Ray-Singer torsion of the manifold. However  the field
theory gives this partition function  as the ratio of a set of
determinants. A standard technique in field theory has been to
define these functional determinants through the analytic
continuation of the zeta functions of the associated
operators.
\par
In this work we investigate the individual determinants that arise
and obtain highly explicit expressions for them. Our
expressions have an intriguing structure of their own
For example, on the lens space $L(2,1)$,  we find that
$$\eqalign{\ln\Det d^*d_{0}&=-{3\zeta(3)\over2\pi^2}+\ln2\cr
\ln\Det d^*d_1&=-{3\zeta(3)\over\pi^2}-2\ln2\cr}\no$$
Similar, though more complicated, expressions occur for each of the lens
spaces $L(p,1)$ for $p=3,4,\dots$. This in turn leads to non-trivial
formulae for $\zeta(3)$: to give two examples we find that
$$\eqalign{\zeta(3)&={2\pi^2\over7}\ln(2)
-{8\over7}\int_0^{\pi/2}dz\,z^2\cot(z)\cr
\zeta(3)&={2\pi^2\over13}\ln 3
-{9\over 13}\int_{0}^{\pi\over3}dz\,z(z+{\pi\over3})\cot(z)
-{9\over 13}\int_{0}^{2\pi\over3}dz\,z(z-{\pi\over3})\cot(z)
\cr}\no$$
these being the formulae that come  from $L(2,1)$ and $L(3,1)$ respectively.
\par
The structure of the paper is as follows. In Section 2  we describe
the topological field theory under consideration. In section 3
we define the Ray-Singer torsion and describe the lens spaces  with which
we work; we also carry
out the non-trivial task of obtaining the
eigenvalues and degeneracies for the Laplacians on $p$-forms acting on these
spaces. Section 4 deals with the lens space $L(2,1)$ ($SO(3)$) and is a
construction of the analytic  continuation of the appropriate $p$-form zeta
functions followed by  a calculation of their associated determinants.
 Sections 5 and 6 describe the analogous
calculation and expressions for the infinite classes of manifolds
corresponding to $L(p,1)$ for $p$ odd and even respectively.
Finally in section 7 we present our conclusions, some comments on the
torsion of $L(p,q)$ for general $q$, and some graphical data
for the resulting determinants and torsion.
\beginsection{Topological Field Theory}
The torsion studied in this paper has its origins in the 1930's, cf.
Franz \ref{1}, where it was combinatorially defined and used to
distinguish various lens spaces from one another. Given a manifold $M$  and a
representation of
its fundamental group $\pi_1(M)$ in a flat bundle $E$, this Reidemeister-Franz
torsion is a real number which is defined as a particular product of ratio's
of volume elements $V^i$ constructed from  the cohomology groups $H^i(M;E)$.
\par
Since volume elements are essentially determinants then, for any alternative
definition of a determinant, an alternative definition of the torsion can
be given. Now if one uses de Rham cohomology to compute $H^i(M;E)$ then
these determinants become determinants of Laplacians $\Delta^E_p$ on $p$-forms
with coefficients in $E$. But zeta functions for elliptic operators can
be used to give finite values to such infinite dimensional determinants and
so an analytic definition of the torsion results and this is the analytic
torsion of Ray and Singer \ref{2,3,4} given in the 1970's;
furthermore this torsion was proved by them to be independent of the
Riemannian metric used to define the Laplacian's  $\Delta^E_p$.
\par
This analytic torsion coincided, for the case of lens spaces, with the
combinatorially defined Reidemeister-Franz torsion. Finally Cheeger and
M\"uller \ref{5,6} independently proved that the analytic Ray-Singer
torsion coincides with the combinatorial Reidemeister-Franz torsion
in all cases.
\par
Infinite dimensional determinants also occur naturally in quantum field
theories when computing correlation functions and partition functions.
In 1978 Schwarz \ref{7} showed how to construct a quantum field theory
on a manifold $M$ whose partition function is a power of the Ray-Singer
torsion on $M$.
\par
Schwarz's construction uses an Abelian gauge theory but in three
dimensions a non-Abelian gauge theory---the $SU(2)$ Chern-Simons
theory---can be constructed and has deep and important properties established
by
Witten in 1988: Its partition
function is the Witten invariant for the three manifold $M$ and the
correlation functions of Wilson loops give the Jones polynomial invariant for
the link determined by the Wilson loops---cf. \ref{8,9}. Finally the weak
coupling limit of the partition function is a  power of the
Ray-Singer torsion.
\par

To define the Ray-Singer torsion, or simply torsion, we take a closed compact
Riemannian
manifold $M$ over which we have  a flat bundle $E$. Let $M$ have a
non-trivial fundamental group  $\pi_1(M)$
which is represented on $E$---this latter property arises very naturally
in the physical gauge theory context where it corresponds simply to the
space of flat connections all of whose content resides in their
holonomy---In any case the torsion is then the real number $T(M,E)$ where
$$\ln T(M,E)=\sum_0^n(-1)^qq\ln \Det \Delta^E_q,\quad n=\dim M\no$$
The metric independence of the torsion requires that we  assume, in the above
definition, that the cohomology ring
$H^*(M;E)$ is trivial; this means that the Laplacians $\Delta^E_q$ have
empty kernels and so are strictly positive definite. Given this fact one
may use zeta functions to define  $ \Det \Delta^E_q$ in the standard way.
Recall that if $P$ is a positive  elliptic differential or
pseudo-differential operator with spectrum
$\{\mu_n\}$ and degeneracies $\{\Gamma_n\}$ then its associated zeta
function $\zeta_P(s)$ is a meromorphic function of $s$, regular at
$s=0$, which is given by
$$\zeta_P(s)=\sum_{\mu_n}{\Gamma_n\over\mu_n^s}\no$$
and its determinant $\Det  P$ is defined by
$$\ln\Det  P=-\left.{d\zeta_P(s)\over ds}\right\vert_{s=0}\no$$
Using this we have
$$\ln T(M,E)=
-\sum_0^n(-1)^qq\left.{d\zeta_{\Delta_q^{E}}(s)\over ds}\right\vert_{s=0}\no$$
\par
Quantum field theories of the type alluded to above
are usually referred to as topological quantum field theories or simply
topological field theories.
\par
It turns out that more than one  topological field theory can be used to
give the torsion, for an excellent review of this question cf.
Birmingham et al. \ref{10}. For example one can take the action
$$S[\omega]=i\int_{M}\omega_{n}d\omega_{n},\qquad \dim M=2n+1\no$$
where $\omega_n$ is an $n$-form. The partition function is then
$$Z[M]=\int {\cal D}\omega\mu[\omega]\exp[S[\omega]] \no$$
$S[\omega] $ has a gauge invariance whereby
$S[\omega]=S[\omega+d\lambda]$ and therefore to define the partition
function it is necessary to integrate over only inequivalent field
configurations. The measure ${\cal D}\omega\mu[\omega]$ thus contains
functional delta functions which constrain the integration and play
the role of gauge fixing, together with their associated
determinants. This measure can be constructed using, for example,
 the Batalin-Vilkovisky BRST construction \ref{11,12}.
\par
We shall be concerned here with the special situation of three dimensions and
with the case where the three manifold $M$ is a lens space.
The topological field theory of interest to us in this paper
is given by the action
$$S[\omega]=i\int_{M}\omega_1 d\omega_1 \no$$
where $\omega_1$ is now a  $1$-form.
To construct the integration measure we will follow the Batalin-Vilkovisky
BRST construction
\ref{11,12}. The essential element of this
construction is what is termed a \lq\lq gauge Fermion'' whose
BRST variation gives the gauge fixing and ghost portion of the BRST
invariant action. Integrating out these fields yields the
contribution $\mu[\omega]$  to the  measure.
\par
The gauge Fermion is constructed by choosing a
gauge fixing for the field $\omega_1$ (which we take to be $\dstar
\omega=0$), and multiplying the condition by an anti-ghost
$c_{\bar{0}}$, which is a $3$-form denoted by its conjugated
Poincar\'e dual label, this indicates its anti-ghost nature also.
Thus the gauge Fermion is given by
$$\Psi  = c_{\specialbar {0}} \dstar \omega_{1}\no$$
The associated BRST variations of these fields are
$$\eqalign{\delta\omega_1 &= -d c_{0},\qquad \delta c_{0}=0\cr
\delta c_{\specialbar {0}}&=i\omega_{\specialbar {0}}
\qquad\delta \omega_{\specialbar {0}}=0\cr
}\no$$
With these definitions it is easy to check that $\delta^2=0$.
The BRST gauge fixed action is then
$${\cal L}= i\omega_1d\omega_1+\delta\Psi $$
which expands to
$${\cal L} = i\omega_1 d \omega_1
- c_{\specialbar {0}}\dstar dc_{0}
+ i\omega_{\specialbar {0}}\dstar \omega_{1}
\no$$
If we integrate out all fields as they appear the resulting partition
function is
$$Z =
(\Det L_{-})^{-{1\over2}}\Det \dstar d_{0}
\no$$
where the operator $L_{-}$ is obtained by integrating out the
$\omega_1$ fields and is a linear operator acting on odd forms.
The partition function is therefore
$$Z={\Det\Delta_{0}\over {\Det\Delta_{1}}^{1\over
4}{\Det\Delta_{3}}^{1\over 4}}$$
Using Poincar\'e duality
the logarithm of this partition function is then given by
$$\ln Z={1\over 4}\left(3\ln\Det \Delta_{0}
-\ln\Det \Delta_{1}\right)$$
and we see it is proportional to  the logarithm of the Ray-Singer torsion.
\par
Our task in what follows is to evaluate the individual components
of this expression both for their usefulness in their own right
and to verify that the combined result agrees with the Ray-Singer
torsion. We do this in the restricted setting  where $M$ belongs to a
class of three dimensional lens spaces.  In the next section we specify the
lens spaces that we work with and obtain the eigenvalues and their degeneracies
of the Laplacians on these spaces.
\beginsection{Lens Spaces}
We now want to turn to field theories defined on lens spaces---for
general background on lens spaces cf. \ref{3,4} and references
therein---briefly, a lens space  can be constructed as follows: Take an
odd dimensional sphere $S^{2n+1}$, considered as a subset of ${\bf C}^n$,
on which a finite cyclic group of rotations $G$, say, acts. The quotient
$S^{2n+1}/G$ of the sphere under this action is a lens space. More
precisely, suppose that $G$ is of order $p$, $(z_1,\ldots,z_n)\in {\bf C}^n$
and the group action takes the form
$$\eqalign{{}&(z_1,\ldots,z_n)\longmapsto (\exp(2\pi i q_1/p)z_1,\ldots,
\exp(2\pi i q_n/p)z_n)\cr
{}&\hbox{with }q_1,\dots,q_n\quad\hbox{integers relatively prime to }p\cr}
\no$$
then  the quotient $S^{2n+1}/G$ is a lens space often denoted by
$L(p;q_1,\dots,q_n)$. A formula
for the torsion of these spaces was first worked out by Ray \ref{2}.
To our knowledge however there is no computation
of the individual determinants of Laplacians on these spaces in the
literature. Since these are of independent field theoretic
significance and from these the torsion is constructed it is
instructive to examine these separately and construct the torsion
from them. This we will proceed to do in the next sections
focusing on the situation that obtains when $n=2$ and $G$ is the
group ${\bf Z}_p\equiv {\bf Z}/p{\bf Z}$. For simplicity we shall denote the
resulting lens space $S^3/{\bf Z}_p=L(p;1,1)$ by $L(p)$, we shall also
denote the lens space $L(p,1,q)$ by $L(p,q)$;
in passing we note that when $p=2$ we have $L(2)={\bf R}P^3\simeq SO(3)$.
\par
The group action above defines a representation $V$, say, of
$\pi_1(L(p))$
and also determines a flat  bundle $F=(V\times S^3)/Z_p$, over $L(p)$. It is
the
determinants of Laplacians and the resulting torsion of this $F$
over $L(p)$ with which we are concerned here. Using
zeta functions the torsion of these lens spaces is therefore given by
$$\ln T(L(p),F)=-\sum_0^3(-1)^qq\left.{d\zeta_{\Delta_q^{F}}(s)\over
ds}\right\vert_{s=0}\no$$
\par
As an aid to the calculation of $\ln T(L(p),F)$ it is useful to introduce the
notation
$$\eqalign{\tau(p,s)&=-\sum_0^3(-1)^qq\zeta_{\Delta_q^{F}}(s)\cr
                 T(p)&=T(L(p),F)\cr}$$
The relationship between the two functions being clearly
$$\ln T(p)=\left.{d \tau(p,s) \over ds}\right\vert_{s=0}\no$$
For $\tau(p,s)$ itself we now have
$$\eqalign{\tau(p,s)&=\zeta_{\Delta_1^{F}}(s)-2\zeta_{\Delta_2^{F}}(s)+
3\zeta_{\Delta_3^{F}}(s)\cr
{}&=3\zeta_{\Delta_0^{F}}(s)-\zeta_{\Delta_1^{F}}(s),\quad
\hbox{using Poincar\'e duality}\cr}\no$$
Combining the standard decomposition $\Delta_p=(\dstar d+d\dstar)_p$,
with the fact that $\ker\dstar\cap\ker d=\emptyset$,
we further obtain the formula
$$\tau(p,s)=2\zeta_{{\dstar d}_0}(s)-\zeta_{{\dstar d}_1}(s)\no$$
We now simplify our notation by labelling
$$\tau_{+}(p,s)=2\zeta_{{\dstar d}_0}(s),\qquad
\tau_{-}(p,s)=\zeta_{{\dstar d}_1}(s)
$$
For the individual zeta functions we denote the eigenvalues and their
degeneracies by $\lambda_n(q,p)$ and $\Gamma_n(q,p)$ respectively giving
the expressions
$$\tau_{+}(p,s)=2\sum_n{\Gamma_n(0,p)\over
\lambda_n^s(0,p)},\qquad
\tau_{-}(p,s)=\sum_n{\Gamma_n(1,p)\over
\lambda_n^s(1,p)}\no$$
It remains to compute these eigenvalues and degeneracies.     The former
are actually independent of $p$ and are fairly easily calculated by the
technique of starting with harmonic forms in ${\bf R}^{2n}$ and
then restricting successively to $S^{2n+1}$ and $L(p)$. In any case they are
given by
$$\eqalign{\lambda_n(0,p)&=n(n+2),\;n=1,2,\dots\cr
           \lambda_n(1,p)&=(n+1)^2,\;n=1,2,\dots\cr}\no$$
\par
To calculate the degeneracies is more difficult;  we make use of the
fact that $S^3$ is a group manifold and proceed as follows: Consider the
Laplacians $\dstar d_q$   on $S^3$, and $\dstar d_q^{F}$ on  $L(p)$
also, if $\lambda$ is an eigenvalue, denote the corresponding
eigenspaces by $\Lambda_q(\lambda)$ and $\Lambda^{F}_q(\lambda)$
respectively. Let
$$v(z)\in \Lambda_q(\lambda),\;\hbox{with }z\in S^3\subset {\bf C}^2,
\qquad \hbox{and }g\in {\bf  Z}_p,\;
\hbox{where }g\equiv\exp[2\pi i j/p],\;0\le j\le (p-1)\no$$
 The element $g$ acts on $v(z)$ to give $g\cdot v(z)$ where
$$\eqalign{g\cdot v(z)&=v(gz)\cr
gz&=(\exp[2\pi i j/p]z_1,\exp[2\pi i j/p]z_2)\cr}\no$$
The above definitions allow us to define the projection $P(\lambda)$ on
$\Lambda_q(\lambda)$ by
$$P(\lambda)v={1\over p}\sum_{g\in {\bf Z}_p}\exp[-2\pi i j/p]g\cdot v\no$$
Evidently
$$[P(\lambda),\dstar d_q]=0\no$$
and  so $P(\lambda)$ projects the space $\Lambda_q(\lambda)$ onto the space
$\Lambda_q^{F}(\lambda)$. Finally this means that we obtain a formula
for the degeneracy $\Gamma_n(q,p)$, namely
$$\eqalign{\Gamma_n(q,p)&=
\tr\left(\left.P\right\vert_{\Lambda_q^{F}(\lambda)}\right)\cr
{}&={1\over p}\sum_{j=0}^{(p-1)}\exp[-2\pi i
j/p]\,\tr\left(\left.g\right\vert_{\Lambda_q^{F}(\lambda)}\right)\cr
}\no$$
To actually apply this formula we now add in the fact that $S^3$ is the
group manifold for $SU(2)$. The Peter--Weyl theorem tells us, in this
case where all representations are self-conjugate, that
$$L^2(S^3)=L^2(SU(2))={\textstyle
\bigoplus\limits_{\mu}}\,\,c_{\mu}D_{\mu}=
{\textstyle \bigoplus\limits_{\mu}}\,\,D_\mu\otimes D_\mu\no$$
where $c_{\mu}$ measures the multiplicity of the representation $\mu$
which must therefore be $\dim D_\mu$.
But Hodge theory gives us the alternative decomposition
$$L^2(S^3)={\textstyle
\bigoplus\limits_{\lambda}}\,\,\Lambda_0(\lambda)\no$$
In addition the Casimir operator for $SU(2)$ is a multiple of the Laplacian
and, if the
 representation label $\mu$ is taken to be the usual half-integer $j$,  then
we know that this Casimir has eigenvalues $j(j+1)$, and also that
$\dim D_j=2j+1$.
These facts identify the Laplacian $\Delta_0=\dstar d_0$ as four times the
Casimir and identify $\Lambda_0(\lambda)$ as $\dim D_j$
copes of $D_j$. Thus if we set $n=2j$, so that $n$ is always integral,
then we have the degeneracy formula
$$\Gamma_n(0,p)={(n+1)\over p}\sum_{j=0}^{(p-1)}\exp[-2\pi i
j/p]\,\chi^{n/2}(2\pi j/p)\no$$
where $\chi^j(\theta)$ denotes the $SU(2)$ character, on $D_j$, for rotation
through the angle $\theta$; i.e.
$$\chi^j(\theta)={\sin((2j+1)\theta)\over\sin(\theta)}\no$$
Hence our explicit degeneracy formula for $0$-forms on $L(p)$ is
$$\Gamma_n(0,p)={(n+1)\over p}\sum_{j=0}^{(p-1)}\exp[-2\pi i
j/p]\,{\sin(2\pi(n+1)j/p)\over\sin(2\pi j/p)}\no$$
\par
We now have to find the analogous formula for the $1$-forms.
The formula that results is
$$\Gamma_n(1,p)={1\over p}\sum_{j=0}^{(p-1)}\exp[-2\pi i
j/p]\left\{n\chi^{(n+1)/2}(2\pi j/p)+(n+2)\chi^{(n-1)/2}(2\pi j/p)\right\}\no$$
or, more explicitly,
$$\Gamma_n(1,p)={1\over p}\sum_{j=0}^{(p-1)}\exp[-2\pi i
j/p]\left\{n{\sin(2\pi(n+2)j/p)\over\sin(2\pi j/p)}+
(n+2){\sin(2\pi nj/p)\over\sin(2\pi j/p)}\right\}\no$$
\par
To simplify the notation we introduce the \lq $p$-averaged character'
$\charav{j}$
which we define by
$$\charav{j}={1\over p}\sum_{j=0}^{(p-1)}\exp[-2\pi i
j/p]\,\chi^j(2 pi j/p)\no$$
Finally this gives us a concrete expression for  $\tau(p,s)$, i.e.
$$\eqalign{\tau(p,s)&=
\sum_n\left\{{2(n+1)\charav{n/2}\over \{n(n+2)\}^{2s}}
-{n\charav{(n+1)/2}+(n+2)\charav{(n-1)/2}\over
(n+1)^{2s}}\right\}\cr
 {}&=\tau_+(p,s)-\tau_-(p,s)\cr
\hbox{where}\quad&\cr
 \tau_+(p,s)&=\sum_n{2(n+1)\charav{n/2}\over \{n(n+2)\}^{s}},
\quad \tau_-(p,s)=\sum_n{n\charav{(n+1)/2}+(n+2)\charav{(n-1)/2}\over
(n+1)^{2s}}\cr}\no$$
\par
To make further progress towards a computation of the
determinants and torsion we need to be able to evaluate these
$p$-averaged characters. This is a somewhat
non-trivial combinatorial task but this task is eased if we use  for
$\chi^j(\theta)$, the
alternative expression
$$\chi^j(\theta)=\sum_{m=-j}^j\exp[2 i m\theta]\no$$
It is also necessary to divide $n$ up into its conjugacy classes mod $p$ by
writing
$$n=pk-j,\;k\in{\bf Z},\, j=0,1,\ldots,(p-1)\no$$
We eventually discover that
\eqlabel{\charformula}
$$\charav{(pk-j)/2}=
\cases{
       \cases{k&for $j=0,2,\ldots,(p-1)$\cr
              k&for $j=1$\cr
              (k-1)&for $j=3,5,\ldots,(p-2)$\cr}
             &if $p$ is odd\cr
           \strut&\null\cr
      \cases{0&for $j=0,2,\dots,(p-2)$ \cr
             2k&for $j=1$\cr
             (2k-1)&for $j=3,5,\ldots,(p-1)$\cr}
             &if $p$ is even\cr}  \no$$

We now lack only one ingredient among those necessary for a calculation of
the determinants and the resulting torsion: this is the construction of
the analytic continuation of the series for $\tau(p,s)$. We shall
construct this in the next section.
The technique we shall use will be more easily followed if we first use it in
a more simple case. Thus, to begin with, we set $p=2$ and then construct the
continuation.

\beginsection{The Analytic Continuation for $p=2$}
The series to be continued are
$$\tau_{+}(p,s)=\sum_{n=1}^{\infty}
{2(n+1)\charav{n/2}\over\{n(n+2)\}^{s}}$$
$$\tau_{-}(p,s)=\sum_{n=1}^{\infty}
{n\charav{(n+1)/2}+(n+2)\charav{(n-1)/2}\over(n+1)^{2s}}$$
and their difference which leads to the torsion
$$\tau(p,s)=\sum_n\left\{{2(n+1)\charav{n/2}\over \{n(n+2)\}^{s}}
-{n\charav{(n+1)/2}+(n+2)\charav{(n-1)/2}\over(n+1)^{2s}}\right\}
 \no$$
These already converges for $\hbox{Re}\, s>3/2$; however a calculation of
the determinants and the torsion requires us to work at $s=0$, hence we
see the need for, and the extent of, the analytic continuation.
\par
Our interest in this section for illustrative purposes is in the
case $p=2$ where we have
$$\eqalign{\tau(2,s)&=\tau_+(2,s)-\tau_-(2,s)\cr
                  {}&=\sum_n\left\{{2(n+1)\left\langle\chi^{n/2}
\right\rangle_2\over \left\{n(n+2)\right\}^s}-
{n\left\langle\chi^{(n+1)/2}\right\rangle_2+
(n+2)\left\langle\chi^{(n-1)/2}\right\rangle_2
\over(n+1)^{2s}}\right\}\cr}\no$$
But using \docref{charformula} we find that
$$\eqalign{\left\langle\chi^{(n+1)/2}\right\rangle_2&=
           \left\langle\chi^{(2k-j+1)/2}\right\rangle_2,\quad (n=2k-j)\cr
        \strut&{}\cr
        {}&=\cases{0,& $j=1$\cr
                    2k+2,& $j=0$\cr}
             \quad\equiv\cases{0,& $n$ odd\cr
                    2k+2,& $n$ even\cr}\cr}\no$$

Similarly
$$\eqalign{\left\langle\chi^{(n-1)/2}\right\rangle_2&=
\left\langle\chi^{(2k-j)/2}\right\rangle_2,\quad (n=2k-j)\cr
        \strut&{}\cr
        {}&=\cases{2k,& $j=1$\cr
                    0,& $j=0$\cr}
             \quad\equiv\cases{(n+1),& $n$ odd\cr
                    0,& $n$ even\cr}\cr}\no$$
Thus $\tau(2,s)$ becomes
$$\eqalign{\tau(2,s)&=\tau_+(2,s)-\tau_-(2,s)\cr
                    {}&=\sum_{n\;{\rm odd}}{
2(n+1)^2\over\left\{n(n+2)\right\}^s}
-\sum_{n\;{\rm even}}{2n(n+2)\over(n+1)^{2s}}\cr}\no$$
Setting $n=(2m-1)$ in $\tau_+(2,s)$ and $n=2m$ in $\tau_-(2,s)$
we have
$$\tau_{+}(2,s)=\sum_{m=1}^\infty{8 m^2\over(4m^2-1)^s},
\qquad\tau_{-}(2,s)=\sum_{m=0}^\infty{4m(2m+2)\over (2m+1)^{2s}}
$$
and
$$
\tau(2,s)=\sum_{m=1}^\infty{8 m^2\over(4m^2-1)^s}
-\sum_{m=0}^\infty{2\over (2m+1)^{(2s-2)}}
+\sum_{m=0}^\infty{2\over (2m+1)^{2s}}\no$$
Now if we use the fact that
$$\sum_{n=1,3,5,\dots}{1\over n^s}=(1-2^{-s})\zeta(s)\no$$
where $\zeta(s)$ is the usual Riemann zeta function then we get
$$\tau_{-}(2,s)=2(1-2^{-(2s-2)})\zeta(2s-2)-2(1-2^{-2s})\zeta(2s)\no$$
and denoting
$$A_{2}(m,0,s)={{(2m)}^2\over{\left\{{(2m)}^2-1\right\}}^s},
\qquad {\rm and}\qquad A_{2}(0,s)=\sum_{m=1}^{\infty}A_{2}(m,0,s)
$$
(This notation is used to agree with the general case to be discussed
in the next section. See also Appendix A.) Thus we have
$$\tau_{+}(2,s)=2A_{2}(0,s)$$
and these combine to give
\eqlabel{\zetaexpress}
$$\tau(2,s)=2A_{2}(0,s)-2(1-2^{-(2s-2)})\zeta(2s-2)+2(1-2^{-2s})\zeta(2s)\no$$
Since the terms involving the Riemann zeta function
already have a well defined continuation it remains to continue
$A_{2}(0,s)$. Now
$$\eqalign{A_{2}(m,0,s)&={4m^2\over(4m^2-1)^s}={4m^2\over (4m^2)^s}
                      \left(1-{1\over4m^2}\right)^{-s}\cr
                {}&={1\over(4m^2)^{(s-
                                    1)}}\left\{1+{s\over4m^2}+\cdots\right\}\cr
                {}&={1\over(4m^2)^{(s-1)}}+{s\over(4m^2)^s}+R(m,s),
                                        \qquad(\hbox{def. of }R(m,s))\cr}\no$$
So that the remainder term $R(m,s)$ is given by
$$\eqalign{R(m,s)&=A_{2}(0,m,s)-{1\over(4m^2)^{(s-1)}} -{s\over(4m^2)^s}\cr
                 &={4m^2\over(4m^2-1)^s}-{1\over(4m^2)^{(s-1)}}-
                                                  {s\over(4m^2)^s}\cr}\no$$
\par
The definition of the remainder term is chosen to ensure that
$$\left\vert R(m,s)\right\vert\le{(\ln m)^\alpha\over m^2}\no$$
and this has the vital consequence that the operations $d/ds$ (at $s=0$) and
$\sum_m$ {\it commute } when applied to $R(m,s)$.
\par
Defining
$$R(s)=\sum_{m=0}^\infty R(m,s)\no$$
allows us to tidy our expressions  up somewhat. Collecting our
regulated expressions we therefore have
\eqlabel{\taupmtwo}
$$\eqalign{\tau_{+}(2,s)&={8\over{4}^{s}}\zeta(2s-2)
+{2s\over 4^s}\zeta(2s)+2 R(s)\qquad {\rm and}\cr
\tau_{-}(2,s)&=2(1-2^{-(2s-2)})\zeta(2s-2)-2(1-2^{-2s})\zeta(2s)\cr}\no$$
In fact the expression for $\tau(2,s)$ can be further tidied up to give
$$\tau(2,s)=2\left\{{8\over4^{(s)}}-1\right\}\zeta(2s-2)
+2\left\{1+{(s-1)\over4^s}\right\}\zeta(2s)+2R(s)\no$$
The series for $R(s)$ is {\it guaranteed} to be convergent and the analytic
continuation is now complete.
\par
Evaluating our expressions at $s=0$ we find
$$\eqalign{\tau_{+}(2,0)&=8\zeta(-2) +2R(0)
\qquad \tau_{-}(2,0)=-6\zeta(-2)\cr
&{\rm and}\qquad
\tau(2,0)=14\zeta(-2)\cr
}\no$$
Observe that with our continuation
$R(0)$ is automatically zero.  Thus noting also that
$\zeta(-2)=0$, we conclude
$$\tau_{+}(2,0)=0\, , \qquad \tau_{-}(2,0)=0\, ,
\qquad and \qquad\tau(2,0)=0\no$$
That $\tau_{\pm}(p,0)=0$ is quite generally true for arbitrary $p$; we
shall see this in the next section and this agrees with general considerations
for  generalised  zeta functions of second order operators on  compact odd
dimensional manifolds.
\par
We can now take the final step which is to
differentiate \docref{taupmtwo} and obtain $\tp(2,0)$ and
$\tm(2,0)$, which we denote by $\tp(2)$,  $\tm(2)$ respectively,
and hence the torsion  $T(2)$.
\par
The resulting expressions  are
$$\tp(2)= 16\zeta '(-2)+2\zeta(0)+2R^\prime(0)$$
$$\tm(2)= -12\zeta '(-2)-2\ln4\zeta(0)$$
and for the torsion
$$\ln T(2)={d\tau(2,0)\over ds}=28\zeta^\prime(-2)
+2(1+\ln4)\zeta(0)+2R^\prime(0)\no$$
But
$$\zeta(0)=-1/2,\qquad\hbox{ and }\qquad \zeta^\prime(-2)=-
{\zeta(3)\over4\pi^2}\quad\hbox{from the functional relation}\no$$
and by our remark above concerning the motive for our choice of definition
for $R(m,s)$ we have
$$\eqalign{R^\prime(0)&=
{d\over ds}\sum_m\left. R(m,s)\right\vert_{s=0}\cr
\Rightarrow R^\prime(0)&=\sum_m\left.{d R(m,s)\over ds}\right\vert_{s=0}\cr
                       &=\sum_m\left[4m^2\left\{\ln(4m^2)-
\ln(4m^2-1)\right\}-1\right]=-
\sum_m\left[4m^2\ln(1-1/4m^2)+1\right]\cr}\no$$
Hence
$$\eqalign{
\tp(2)&=-{4\over\pi^2}\zeta(3)-1
-2\sum_m\left[4m^2\ln(1-1/4m^2)+1\right]\cr
\tm(2)&={3\over\pi^2}\zeta(3)+2\ln2\qquad {\rm and}\cr
\ln T(2)&=-{7\over\pi^2}\zeta(3)-1-2\ln(2)-
2\sum_m\left[4m^2\ln(1-1/4m^2)+1\right]}\no$$
However the series for $R^\prime(0)$ can be expressed as a trigonometric
integral; in fact, as a special case of more general results which will be
derived below, we have
$$\sum_{m=1}^\infty\left[4m^2\ln(1-1/4m^2)+1\right]=-{1\over2}+
{4\over\pi^2}\int_0^{\pi/2}dz\,z^2\cot(z)\no$$
which means that
\eqlabel{\tortwoexpr}
$$\eqalign{\tp(2)&=-{4\over \pi^2}\zeta(3)
-{8\over\pi^2}\int_{0}^{\pi\over2}dz z^2\cot(z)\cr
\tm(2)&={3\over\pi^2}\zeta(3)+2\ln2\qquad {\rm and}\cr
\ln T(2)&=-{7\over\pi^2}\zeta(3)-2\ln(2)-
{8\over\pi^2}\int_0^{\pi/2}dz\,z^2\cot(z)}\no$$
The formula \docref{tortwoexpr} above for $T(2)$ can be pushed even further:
By using it with Ray's expression \ref{2} for the torsion  we can
deduce that
\eqlabel{\rayexpr}
$$\eqalign{\ln T(p)&=-{4\over p}\sum_{j=1}^{(p-1)}\sum_{k=1}^p\cos({2jk\pi\over
p})\ln(2\sin({2k\pi\over p}))\exp[{2k\pi i\over  p}]\cr
&=-4\ln\left[2 \sin({\pi\over p})\right]\cr}\no$$
which, for $p=2$, becomes simply
$$\ln T(2)=-4\ln(2)\no$$
Hence we straightaway  have the identity
$$-4\ln(2)=-{7\over\pi^2}\zeta(3)-2\ln(2)-
{8\over\pi^2}\int_0^{\pi/2}dz\,z^2\cot(z)\no$$
Or
\eqlabel{\zetaptwo}
$$\zeta(3)={2\pi^2\over7}\ln(2)-{8\over7}\int_0^{\pi/2}dz\,z^2\cot(z)\no$$
In other words our computation of the torsion has given us a formula for
the zeta function at its first odd argument. Equivalently we can use
this relation to eliminate the integral and obtain quite simple
expressions for  the
logarithms of the determinants of the Laplacians on zero and one
forms respectively.  We conclude this section by quoting these
results
\par
Noting first of all  that the expressions for  $\tp(2)$ and
$\tm(2)$ in their simplest form now become
\eqlabel{\tpmp}
$$\tp(2)= {3\over \pi^2}\zeta(3)-2\ln 2\qquad {\rm and }
\qquad\tm(2)={3\over \pi^2}\zeta(3)+2\ln 2\no$$
and from the definitions of these objects we have at once that
\eqlabel{\Detsptwo}
$$\eqalign{
\ln \Det d^{*}d_{0} &= -{3\over 2\pi^2}\zeta(3) +\ln2\cr
\ln \Det d^{*}d_{1} &= -{3\over \pi^2}\zeta(3)-2 \ln2\cr
}\no$$
It is interesting to note the role that the Riemann zeta function
$\zeta(3)$ plays in these expressions. Since these are expressions
for volume elements on the  discrete moduli spaces
  associated
with the Laplacians, we expect that there are deeper things to be learned from
a further study of such expressions.
\par
In the next section we tackle the continuation for arbitrary $p$.
\beginsection{The Determinants and the Torsion for $p$ Odd.}
The analytic continuation for a general value of $p$ naturally divides into
two cases: $p$ odd and $p$ even; in fact we shall see below that the case for
$p$ even further divides into two subcases which correspond to
$p=0,2\,{\rm mod\,}4$. Due to the size of the expressions it
is now much more convenient to continue $\tau_+(p,s)$ and $\tau_-(p,s)$
separately and then combine them into  expressions for the torsion.
We deal first with $\tau_+(p,s)$  and begin with $p$ odd.
\subsec{The Function $\tau_{+}(p,s)$ and $\ln\Det\dstar d_0$}
Let us  recall that  $\tau_+(p,s)$ is given by
$$\tau_+(p,s)=\sum_{n=1}^\infty
{2(n+1)\left\langle\chi^{n/2}
\right\rangle_p\over \left\{n(n+2)\right\}^s}
\no$$
\par
Reference to the general character formula \docref{charformula} shows that we
must resolve $n$ into its conjugacy classes  mod $p$ by writing
$$n=pk-j,\;\,j=0,\ldots,(p-1)\no$$
and that we must distinguish the two parities  of $j$. To implement these
requirements we set
\eqlabel{\paraties}
$$p=2r+1,\qquad\hbox{and parametrise }
\quad\cases{j& odd by $j=2l+1$, $l=0,1,\ldots,(r-1)$\cr
            j& even by $j=2l$, $l=0,1,\ldots,r$\cr}\no$$
This gives
$$\eqalign{\tau_+(p,s)&=2\sum_{j=0}^{p-1}\sum_{k=1}^\infty
{(pk-j+1)\left\langle\chi^{(pk-j)/2}\right\rangle_p
                          \over\left\{(pk-j)(pk-j+2)\right\}^s}\cr
                       &=2\sum_{k=1}^\infty
                      \left[\sum_{l=0}^{r-1}{(pk-2l)
                         \left\langle\chi^{(pk-2l-1)/2}\right\rangle_p
                          \over\left\{(pk-2l-1)(pk-2l+1)\right\}^s}
                        +\sum_{l=0}^r{(pk-2l+1)
                         \left\langle\chi^{(pk-2l)/2}\right\rangle_p
                       \over\left\{(pk-2l)(pk-2l+2)\right\}^s}\right]\cr}\no$$
Then when we use \docref{charformula} for $p$ odd we get
\eqlabel{\tauplusodd}
$$\eqalign{\tau_+(p,s)
=\sum_{k=1}^\infty\left[{2pk^2\over\left\{(pk-1)(pk+1)\right\}^s}\right.
              & +\sum_{l=1}^{r-1}{2(k-1)(pk-2l)\over\left\{(pk-2l-1)(pk-
                                                       2l+1)\right\}^s}\cr
             &+\sum_{l=0}^r\left.{2k(pk-2l+1)\over\left\{(pk-2l)(pk-
                                            2l+2)\right\}^s}\right]\cr}\no$$
To aid in marshalling  the combinatorics of $\tau_+(p,s)$ we define
$H_{p}(k,s,x)$ by
$$H_{p}(k,s,x)={pk(pk+x)
\over\left\{(pk+x-1)(pk+x+1)\right\}^s}\no$$
The point being that each of the three summands in \docref{tauplusodd} is
of the form $H_{p}(k,s,\lambda)$ for appropriate $\lambda$. To see this we
introduce precisely $p$ constants of the type $\lambda$ defined by
$$\cases{\lambda_0=0&\cr
         \lambda_l=-2l+1,\;\,l=0,\ldots,r&\cr
          \bar\lambda_l=-2l+p,\;\,l=1,\ldots,(r-1)&\cr}\no$$
With this notation it can be checked that $\tau_+(p,s)$ is given by
$$\tau_+(p,s)={2\over p}\sum_{k=1}^\infty\left[H_{p}(k,s,\lambda_0)
                          +\sum_{l=1}^{(r-1)}H_{p}(k,s,\bar\lambda_l)
                           +\sum_{l=0}^{r}H_{p}(k,s,\lambda_l)\right]\no$$
Also if we denote the entire set of $\lambda$'s by $\{\lambda\}$
i.e.
$$\{\lambda\}\equiv\{\lambda_0,\lambda_l,\bar\lambda_l\}=\{-(p-2),
\ldots,-5,-3,-1,0,1,3,5\ldots(p-2)\}\no$$
then we have the even more concise expression
\eqlabel{\tauHse}
$$\tau_+(p,s)={2\over p}\sum_{\{\lambda\}}\sum_{k=1}^\infty
H_{p}(k,\lambda,s)\no$$
The functions $H_{p}(\lambda,s)$ obtained by summing over $k$ then
form a set of $p$ functions whose derivative at $s=0$ can be viewed as
living on the appropriate space of sections for the Laplacian acting on $0$
forms on the lens space $L(p)$; taking the  trace over these functions viewed
as forming a matrix then gives the analytic continuation of the
determinant of the Laplacian.
\par
Next observing
$$H_{p}(k,x,s)
={{(pk+x)}^2\over\{{(pk+x)}^2-1\}^s}
-x{(pk+x)\over\{{(pk+x)}^2-1\}^s}
\no$$
we see that there are therefore two additional functions of interest here
i.e.
$$A_{p}(k,x,s)={{(pk+x)}^2\over{\left\{{(pk+x)}^2-1\right\}}^s}\no$$
and
$$B_{p}(k,x,s)={(pk+x)\over{\left\{{(pk+x)}^2-1\right\}}^s}\no$$
and we have
$$H_{p}(k,x,s)=A_{p}(k,x,s)-xB_{p}(k,x,s)\no$$
and we can equally write
\eqlabel{\taupab}
$$\tau_{+}(p,s)
={2\over p}\sum_{\{\lambda\}}\left[A_{p}(\lambda,s)
-\lambda B_{p}(\lambda,s)\right]\no$$
Let us further note that in the set $\{\lambda\}$ the non zero
elements come in pairs of the form $\{\lambda,-\lambda\}$.
Thus we can further write
$$\tau_{+}(p,s)={2\over p}A_{p}(0,s)+{2\over p}\sum_{l=1}^{r-1}
\left[A_{p}^{+}(2l+1,s)-(2l+1)B_{p}^{-}(2l+1,s)\right]$$
where the $\mp$  superscripts refer to the symmetric
 or  anti-symmetric combination with respect to
the first argument: i.e.
$A_{p}^{+}(x,s)=A_{p}(x,s)+A_{p}(-x,s)$
and $B_{p}^{-}(x,s)=B_{p}(x,s)-B_{p}(-x,s)$.
We relegate the details of the computation of these functions
and their analytic continuation to appendices A and B---the calculations are
generalisations of those performed for $p=2$. Quoting here from appendices A
and B we have that the relevant functions  and their
derivatives at $s=0$ are given by
$$\eqalign{A_{p}(x,0)&=p^2\zeta(-2,1+{x\over p})
= -{x(x+p)(2x+p)\over 6p}\cr
B_{p}(x,0)&=p\zeta(-1,1+{x\over p})+{1\over 2p}=-{p\over
12}-{x\over p}+{x^2+1\over 2 p}\cr
H_{p}(x,0)&=-{px\over 12}-{x\over 2p}+{x^3\over 6p}\cr
}\no$$
Which immediately implies that $H_{p}^{+}(x,0)=0$ and we have our
first result that for $p$ odd
$$\tau_{+}(p,0)=0$$
The significance of this is that the analytic continuation of the
scaling dimension of the determinant  is zero.
\par
Next we note again  from appendices A and B that
\eqlabel{\Aplusp}
$$\eqalign{A_{p}^{+}(x)
=-{p^2\over\pi^2}\zeta(3)
&-{p^2\over\pi^2}\int_{0}^{(x+1)\pi\over p}
dz\, z(z-{2\pi x\over p})\cot(z)
-{p^2\over\pi^2}\int_{0}^{(x-1)\pi\over p}
dz\, z(z-{2\pi x\over p})\cot(z)\cr
&-x^2\ln\left[{2\sin\left({(x+1)\pi/ p}\right)\over x+1}\right]
-x^2\ln\left[{2\sin\left({(x-1)\pi/ p}\right)\over x-1}\right]\cr
}\no$$
and
\eqlabel{\Bminusp}
$$\eqalign{B_{p}^{-}(x)
&= {p\over\pi}\int_{0}^{(x+1)\pi\over p}dz\,z\cot(z)
+{p\over\pi}\int_{0}^{(x-1)\pi\over p}dz\, z\cot(z)\cr
&\qquad-x\ln\left[{2\sin\left({(x+1)\pi/ p}\right)\over x+1}\right]
-x\ln\left[{2\sin\left({(x-1)\pi/ p}\right)\over x-1}\right]\cr
}\no$$

Now  combining our expressions \docref{Aplusp}, \docref{Bminusp}  and
 summing over $\{\lambda\}$ we obtain
for $\tp(p)$ the expression
\eqlabel{\taupp}
$$\eqalign{\tp(p)
&=-{2\over p}\left[{p^3\over2\pi^2}\zeta(3)
+{p^2\over\pi^2}\int_{0}^{\pi\over p}dz\, z^2\cot(z)
+{p^2\over\pi^2}\int_{0}^{(p-1)\pi\over p}
dz\, z(z-{(p-2)\pi\over p})\cot(z)\right.\cr
&\qquad\left.+2\sum_{l=1}^{(p-3)\over2}
{p^2\over\pi^2}\int_{0}^{2l\pi\over p}dz\, z(z-{2l\pi\over p})\cot(z)
\right]\cr
}\no$$
for $p$ odd. But the expression $\tp(p)$ above is $2\ln\Det\dstar d_0 $:
i.e. twice the logarithm  of the
Laplacian on 0-forms for the lens space $L(p)$.  More precisely our
analytic continuation has shown us that
$$\eqalign{\ln\Det\dstar d_0
&={1\over p}\left[{p^3\over2\pi^2}\zeta(3)
+{p^2\over\pi^2}\int_{0}^{\pi\over p}dz z^2\cot(z)
+{p^2\over\pi^2}\int_{0}^{(p-1)\pi\over p}dz
z(z-{(p-2)\pi\over p})\cot(z)\right.\cr
&\qquad\left.+2\sum_{l=1}^{(p-3)\over2}
{p^2\over\pi^2}\int_{0}^{2l\pi\over p}dz z(z-{2l\pi\over p})\cot(z)
\right]\cr
}\no$$
\subsec{The Function $\tau_{-}(p,s)$ and $\ln\Det\dstar d_1$}
Let us now turn to $\tau_{-}(p,s)$ for $p$ odd.
$$\tau_{-}(p,s)=\sum_{n=1}^{\infty}
{n\left<\chi^{(n+1)\over2}\right>_{p}
+(n+2)\left<\chi^{(n-1)\over2}\right>_{p}\over {(n+1)}^{2s}}\no$$
which on decomposing $n$ over the conjugacy classes mod $p$
$n=pk-j,\quad j=0,\dots,(p-1)$
as for $\tau_{+}(p,s)$, distinguishing the two parities as in
\docref{paraties} yields
\vfill
$$\eqalign{\tau_{-}(p,s)
&=\sum_{k=1}^{\infty}{kp{\left\langle\chi^{kp+1\over2}\right\rangle}_{p}
+(kp+2){\left\langle\chi^{kp-1\over2}\right\rangle}_{p}\over{(kp+1)}^{2s}}
+\sum_{k=1}^{\infty}{(kp-2){\left\langle\chi^{kp-1\over2}\right\rangle}_{p}
+kp{\left\langle\chi^{kp-3\over2}\right\rangle}_{p}\over{(kp-1)}^{2s}}\cr
&\qquad+\sum_{k=1}^{\infty}\sum_{l=2}^{(p-3)\over2}
{(kp-2l){\left\langle\chi^{kp-2l+1\over2}\right\rangle}_{p}
+(kp-2l+2){\left\langle\chi^{kp-2l-1\over2}\right\rangle}_{p}\over{(kp-2l+1)}^{2s}}\cr
&\qquad+\sum_{k=1}^{\infty}{(kp-p+1){\left\langle\chi^{kp-p\over2}\right\rangle}_{p}
+(kp-p+3){\left\langle\chi^{kp-p-2\over2}\right\rangle}_{p}\over{(kp-p+2)}^{2s}}\cr
&\qquad+\sum_{k=1}^{\infty}\sum_{l=0}^{r-1}
{(kp-2l-1){\left\langle\chi^{kp-2l\over2}\right\rangle}_{p}
+(kp-2l+1){\left\langle\chi^{kp-2l-2\over2}\right\rangle}_{p}\over{(kp-2l)}^{2s}}\cr
}\no$$
Using our degeneracy
formula \docref{charformula} we have with a little re-arrangement
$$\eqalign{\tau_{-}(p,s)
&=\sum_{k=1}^{\infty}{2(kp+1)k+kp\over{(kp+1)}^{2s}}
+\sum_{k=1}^{\infty}{2(kp-1)k-kp\over{(kp-1)}^{2s}}\cr
&\qquad+\sum_{l=-{(p-3)\over2}}^{(p-3)\over2}
\sum_{k=1}^{\infty}{2(kp+2l)k\over{(kp+2l)}^{2s}}\cr
}\no$$
We now observe that this is of the form
\eqlabel{\taumser}
$$\eqalign{\tau_{-}(p,s)
&={2\over p}\sum_{\{\nu\}}\sum_{k=1}^{\infty}
{kp(kp+\nu)\over{(kp+\nu)}^{2s}}\cr
&+\sum_{k=1}^{\infty}{kp\over{(kp+1)}^{2s}}
-\sum_{k=1}^{\infty}{kp\over{(kp-1)}^{2s}}\cr
}\no$$
where we have denoted by $\{\nu\}$ the set
$$\{\nu\}\equiv\{\nu_0,\nu_l,\bar\nu_l\}=\{-(p-3),
\ldots,-6,-4,-2,-1,0,1,2,4,6\ldots(p-3)\}\no$$
The sums occuring in \docref{taumser} are naturally expressible in terms of
Hurwitz zeta functions and this provides the natural analytic
continuation of this expression. We also note that
$$\sum_{\{\nu\}}\sum_{k=1}^{\infty}{1\over {(kp+\nu)}^{(2s-2)}}
=\zeta(2s-2)-1-\sum_{l=1}^{(p-3)\over2}{(2l)}^{(2-2s)}\no$$
On utilising these relations we find that
$$\eqalign{\tau_{-}(p,s)
&={2\over p}
\left[\zeta(2s-2)-1-\sum_{l=1}^{(p-3)\over2}{(2l)}^{2-2s}\right]\cr
&\qquad (p-2){p}^{-2s}
\left[\zeta(2s-1,1+{1\over p})-\zeta(2s-1,1-{1\over p})\right]\cr
&\qquad-2{p}^{-2s}\sum_{l=-{(p-3)\over2}}^{(p-3)\over2}
(2l)\zeta(2s-1,1+{2l\over p})\cr
&\qquad-{p}^{-2s}
\left[\zeta(2s,1+{1\over p})+\zeta(2s,1-{1\over p})\right]
}\no$$
If we analytically continue the RHS of this expression to $s=0$,
and  observe that $\zeta(-1,1+x)-\zeta(-1,1-x)=-x$, then we obtain
$$\tau_{-}(p,0)=-{2\over p}
\left[1+\sum_{l=1}^{(p-3)\over2}{(2l)}^2\right]
+(p-2)(-{1\over p})
-2\sum_{l=1}^{(p-3)\over2}(2l)(-{2l\over p})\no$$
which  immediately gives
$$\tau_{-}(p,0)=0\no$$
Thus again the scaling dimension of the associated determinant is
zero.
\par
Passing to $\tm(p)$ by taking the derivative at $s=0$ gives
$$\eqalign{\tm(p)
&={4\over p}\left[\zeta '(-2)-\ln{p}
+\sum_{l=1}^{(p-3)\over2}{(2l)}^2\ln[{2l\over p}]\right]\cr
&\qquad+2(p-2)\left[\zeta '(-1,1+{1\over p})
-\zeta '(-1,1-{1\over p})\right]\cr
&\qquad-4\sum_{l=1}^{(p-3)\over 2}(2l)
\left[\zeta '(-1,1+{2l\over p})-\zeta '(-1,1-{2l\over p})\right]\cr
&\qquad-2
\left[\zeta '(0,1+{1\over p})+\zeta '(0,1-{1\over p})\right]\cr
}\no$$
A useful identity for Hurwitz zeta functions derived in
appendix C is
$$\zeta '(-1,1+x) -\zeta '(-1,1-x)=
-x\ln\left[{2\sin(\pi x)\over x}\right]+{1\over \pi}\int_{0}^{\pi
x}dz z\cot(z)$$
Using this and the expression
$\zeta '(-2)=-{\zeta(3)/ 4{\pi}^2}$
we conclude that
\eqlabel{\taump}
$$\eqalign{\tm(p)
&= -{1\over p{\pi}^2}\zeta(3)
+{4\over p}\ln\left[2\sin({\pi\over p})\right]
+{4\over p}\sum_{l=1}^{(p-3)\over 2}
{(2l)}^2\ln\left[2\sin({2\pi l\over p})\right]\cr
&\qquad+{2(p-2)\over \pi}\int_{0}^{\pi\over p}dz z\cot(z)
-{4\over\pi}\sum_{l=1}^{(p-3)\over 2}
(2l)\int_{0}^{2\pi l\over p}dz z\cot(z)\cr
}\no$$
 But $-\tm(p)$ is the analytic continuation which gives
$\ln \Det d^*d_{1}$. Hence we find that
$$\eqalign{\ln \Det d^*d_{1}
&= {1\over p{\pi}^2}\zeta(3)
+{4\over p}\ln\left[2\sin({\pi\over p})\right]
+{4\over p}\sum_{l=1}^{(p-3)\over 2}
{(2l)}^2\ln\left[2\sin({2\pi l\over p})\right]\cr
&\qquad+{2(p-2)\over \pi}\int_{0}^{\pi\over p}dz z\cot(z)
-{4\over\pi}\sum_{l=1}^{(p-3)\over 2}
(2l)\int_{0}^{2\pi l\over p}dz z\cot(z)\cr
}\no$$
\subsec{The Torsion $T(p)$ and $\tau(p,s)$.}
We are therefore now in a position to put these together and
obtain an expression for the torsion.
The first observation is  that since $\tau_{+}(p,0)$ and
$\tau_{-}(p,0)$ are both zero we have
$$\tau(p,0)=0\no$$
This vanishing of $\tau(p,0)$ is related to the metric independence of the
torsion something which has  been established  quite generally by
Ray and Singer \ref{3}.
\par
The torsion $T(p)$ itself is given by the difference of
\docref{taupp} and \docref{taump}. Combining these two
expressions we find,
upon a little  simplification, that   $T(p)$ is determined by the
equation
\eqlabel{\ourexpr}
$$\eqalign{&\ln T(p)
=-{2\over p}\left[{(p^3-1)\over2\pi^2}\zeta(3)
+2\ln\left[2\sin({\pi\over p})\right]
+2\sum_{l=1}^{(p-3)\over 2}
4l^2\ln\left[2\sin({2l\pi\over p})\right] \right.\cr
&\qquad+{p^2\over\pi^2}\int_{0}^{\pi\over p}dz\,
z(z+{(p-2)\pi\over p})\cot(z)
+{p^2\over\pi^2}\int_{0}^{(p-1)\pi\over p}dz\,
z(z-{(p-2)\pi\over p})\cot(z)\cr
&\qquad\left.+2\sum_{l=1}^{(p-3)\over 2}
{p^2\over\pi^2}\int_{0}^{2l\pi\over p}dz\, z(z-{4l\pi\over p})\cot(z)
\right]\cr
}\no$$
Now the expression for the torsion from Ray's calculation
gave the alternative expression
\docref{rayexpr} i.e.
$$\ln T(p) = -4\ln\left[2 \sin({\pi\over p})\right]\no$$
We have verified that these two expressions \docref{ourexpr} and
\docref{rayexpr} agree  numerically, yet
it is not transparent by inspection that this should be so; also using
C.41 of appendix C we can reduce \docref{ourexpr} to Ray's expression. One may
conclude that,
by following two alternate derivations, we have arrived at what is a
sequence of non-trivial identities. As we
saw in the case of $p=2$ utilising these identities  one can
obtain non-trivial formulae for $\zeta(3)$ and also can
be used to further simplify the expressions for the individual
determinants.
The resulting expressions for $\zeta(3)$ are
$$\eqalign{\zeta(3)&={2\pi^2\over (p^3-1)}\left[
2(p-1)\ln\left[2\sin({\pi\over p})\right]
-2\sum_{l=1}^{(p-3)\over 2}
4l^2\ln\left[2\sin({2l\pi\over p})\right] \right.\cr
&\qquad-{p^2\over\pi^2}\int_{0}^{\pi\over p}dz\,
z(z+{(p-2)\pi\over p})\cot(z)
-{p^2\over\pi^2}\int_{0}^{(p-1)\pi\over p}dz\,
z(z-{(p-2)\pi\over p})\cot(z)\cr
&\qquad\left.-2\sum_{l=1}^{(p-3)\over 2}
{p^2\over\pi^2}\int_{0}^{2l\pi\over p}dz\, z(z-{4l\pi\over p})\cot(z)
\right]\cr
}\no$$
\par
For the sake of illustration let us quote the implications of these
formulae for the simplest  odd case: $p=3$. On utilising all of the
information at our disposal we find that
$$\eqalign{\tp(3)&=-{\zeta(3)\over 3\pi^2}-{4\over3}\ln3
+{2\over\pi}\int_{0}^{\pi\over3}dz\,z \cot(z)\cr
\tm(3)&=-{\zeta(3)\over 3\pi^2}+{2\over3}\ln3
+{2\over\pi}\int_{0}^{\pi\over3}dz\,z \cot(z)\qquad {\rm and}\cr
\ln T(3)&=-2\ln3\cr
}$$
The relation between the expressions we have obtained which we
expressed in terms of a formula for $\zeta(3)$ (the analog of
\docref{zetaptwo} for $p=2$) becomes
$$\zeta(3)={2\pi^2\over13}\ln 3
-{9\over 13}\int_{0}^{\pi\over3}dz\,z(z+{\pi\over3})\cot(z)
-{9\over 13}\int_{0}^{2\pi\over3}dz\,z(z-{\pi\over3})\cot(z)$$
We will now turn to the case of even $p$.
\beginsection{Determinants and the Torsion for $p$ Even.}
When $p$ is even we follow a
slightly different route to that used in the previous section but we  arrive
at  expressions of a similar general form for the respective
determinants and their corresponding torsion.
\subsec{The Function $\tau_{+}(p,s)$ and $\ln\Det\dstar d_0$}
Let us first obtain series expressions for $\tau_{+}(p,s)$ for $p$
even, and observe that the same functions as those encountered for $p$
odd enter these also.
Recalling our expression for $\tau_{+}(p,s)$
$$\tau_{+}(p,s)=
\sum_{n=1}^{\infty}
{2(n+1)\left<\chi^{n\over2}\right>_{p}\over{\left\{{(n+1)}^2-1\right\}}^s}
\no$$
We again resolve $n$ into its conjugacy classes  mod $p$ by writing
$$n=pk-j,\;\,j=0,\ldots,(p-1)\no$$ and distinguish
the two parities  of $j$ by  setting
$$p=2r,\qquad\hbox{and parametrise }
\quad\cases{j& odd by $j=2l+1$, $l=0,1,\ldots,(r-1)$\cr
            j& even by $j=2l$, $l=0,1,\ldots,(r-1)$\cr}\no$$
Reference to the general character formula \docref{charformula}
shows that only $j$ odd contributes
and  we obtain
$$\tau_{+}(p,s)=\sum_{k=1}^{\infty}{4p k\over{\left\{{(pk)}^2-1\right\}}^s}
+\sum_{l=0}^{(p-2)\over2}
\sum_{k=1}^{\infty}
{2(pk-2l)(2k-1)\over{\left\{{(pk-2l)}^2-1\right\}}^s}
\no$$
which gives
\eqlabel{\taupevens}
$$
\tau_{+}(p,s)={4\over p}\sum_{k=1}^{\infty}\left[\sum_{l=0}^{(p-2)\over2}
{{(pk-2l)}^2\over{\left\{{(pk-2l)}^2-1\right\}}^s}
+\sum_{l=1}^{(p-2)\over2}
(2l-{p\over2}){(pk-2l)\over{\left\{{(pk-2l)}^2-1\right\}}^s}\right]
\no$$
We recognise the expressions arising as the
functions from the preceding analysis in the case
of $p$ odd and which are analysed in appendices A and B. We can
therefore write \docref{taupevens} as
\eqlabel{\taupseone}
$$
\tau_{+}(p,s)={4\over p}\sum_{l=0}^{r-1}
\left[A_{p}(-2l,s)+(2l-r)B_{p}(-2l,s)\right]
\no$$
Some further rearrangement will allow us to
write these again in terms of the
symmetric and anti-symmetric parts of $A_{p}(x,s)$ and $B_{p}(x,s)$
respectively.
Note first of all, however that the term involving
$A_{p}$ is a sum over all even conjugacy classes
i.e.
$$\eqalign{\sum_{l=1}^{r-1}A(-2l,s)
&=\sum_{l=1}^{r-1}\sum_{k=1}^{\infty}
{{(2rk-2l)}^2\over{\left\{{(2rk-2l)}^2-1\right\}}^{s}}\cr
&=\sum_{m=1}^{\infty}{{(2m)}^2\over{\left\{{(2m)}^2-1\right\}}^{s}}\cr
}\no$$
and is therefore ${\tau_{+}(2,s)/2}$;
our expression \docref{taupseone} for $\tau_{+}(p,s)$ can
hence be written as
$$
\tau_{+}(p,s)={2\over p}\left[\tau_{+}(2,s)
+2\sum_{l=1}^{r-1}(2l-r)B_{p}(-2l,s)\right]
\no$$
Now when $l$ ranges from $1$ to $r-1$, $(2l-r)$ ranges over the set
$$-(r-2),(r-4),\dots,(r-4),(r-2)\no$$
This allows us to divide the range up into a sum from $1$ up to the integer
part of ${(r-1)/ 2}$ which we denote by
$\left[{(r-1)/2}\right]$
Thus
$$\tau_{+}(p,s)={2\over p}\left[\tau_{+}(2,s)
+2\sum_{l=1}^{\left[{(r-1)\over2}\right]}(r-2l)
\left[B_{p}(-p+2l,s)-B_{p}(-2l,s)\right]\right]
\no$$
Observing that
$$B_{p}(-p+x,s)={x\over{\left\{{x}^2-1\right\}}^s}+B_{p}(x,s)$$
This gives
\eqlabel{\taupsetwo}
$$ \tau_{+}(p,s)={2\over p}\left[\tau_{+}(2,s)
+2\sum_{l=1}^{\left[{(p-2)\over4}\right]}({p\over2}-2l)
\left[B_{p}^{-}(2l,s)
+{2l\over{\left\{{x}^2-1\right\}}^s}\right]\right]
\no$$
Noting that $\tau_{+}(2,0)=0$ and that $B_{p}^{-}(x)=-x$
we see again immediately that
$$\tau_{+}(p,0)=0\no$$
Again as we expect the scaling dimension of the associated
determinant is  zero.

Proceeding now to the expression for the determinant itself, we find
the resulting expression from \docref{taupsetwo}
for the derivative at $s=0$ is
$$\tau_{p}'(p,0)={2\over p}\left[\tp(2)
+\sum_{l=1}^{\left[{(p-2)\over4}\right]}(p-4l)
\left[B_{p}^{-}(2l)
-2l\ln[{2l}^2-1]\right]\right]
\no$$
Substituting for $B_{p}^{-}(2l)$ from \docref{Bminusp}  gives
\eqlabel{\taupfin}
$$\eqalign{\tp(p)&={2\over p}\left[\tp(2)
+{p\over\pi}\sum_{l=1}^{\left[{(p-2)\over4}\right]}(p-4l)
\left[\int_{0}^{(2l+1)\pi\over p}z\cot(z)
+\int_{0}^{(2l-1)\pi\over p}dz z\cot(z)\right.\right.\cr
&\qquad\left.\left.
-{2l\pi\over p}\ln\left[2\sin\left({(2l+1)\pi\over p}\right)\right]
-{2l\pi\over p}\ln\left[2\sin\left({(2l-1)\pi\over p}\right)\right]
\right]\right]\cr
}\no$$
where from \docref{tpmp} $\tp(2)={3\over\pi^2}\zeta(3)-2\ln2$ .
As we see the case of even $p$ divides naturally into two classes
$p=2\, {\rm mod\,}4$ and $p=0\, {\rm mod\,}4$.
Making this division we can further simplify things to obtain
$$\eqalign{\tp(p)
&={2\over r}\left[{3\over 2\pi^2}\zeta(3)
-2\sum_{l=1}^{(r-3)\over
2}\left[(2l+1)(r-2l-1)-1\right]\ln\left[2\sin\left({(2l+1)\pi\over
2r}\right)\right]\right.\cr
&\left.
+{2r(r-2)\over\pi}\int_{0}^{\pi\over 2r}dz\,z\cot(z)
+{4r\over\pi}\sum_{l=1}^{(r-2)\over2}(r-2l-1)
\int_{0}^{(2l+1)\pi\over 2r}dz\,z\cot(z)\right]\; p=2\mod4\cr
}$$
and
$$\eqalign{\tp(p)
&={2\over r}\left[{3\over 2\pi^2}\zeta(3)-\ln2
-2\sum_{l=1}^{(r-2)\over
2}\left[(2l+1)(r-2l-1)-1\right]\ln\left[2\sin\left({(2l+1)\pi\over
2r}\right)\right]\right.\cr
&\left.
+{2r(r-2)\over\pi}\int_{0}^{\pi\over 2r}dz\,z\cot(z)
+{4r\over\pi}\sum_{l=1}^{(r-2)\over2}(r-2l)
\int_{0}^{(2l+1)\pi\over 2r}dz\,z\cot(z)\right]\;p=0\mod4\cr
}$$
We note that these expressions agree with the result for
$p=2$ and, for $p=4$, we note in passing that inspection of the series
shows that $\tau_{\pm}'(4,0)={\tau_{\pm}'(2,0)/2}$; this turns out to be
also a property of (the logarithm of) the torsion itself, i.e. $T(p)$
satisifies $\ln T(4)=(\ln T(2))/2$.
\par
We therefore have from \docref{taupfin} an expression for the
appropriate logarithmic determinant on the lens space $L(p)$ namely
$$\eqalign{\ln\Det \dstar d_0
&= -{3\over p\pi^2}\zeta(3)+{2\over p}\ln2\cr
&\quad-{2\over p}
\sum_{l=1}^{\left[{(p-2)\over4}\right]}({p\over2}-2l)
\left[{p\over\pi}\int_{0}^{(2l+1)\pi\over p}z\cot(z)
+{p\over\pi}\int_{0}^{(2l-1)\over p}dz z\cot(z)\right.\cr
&\quad\left.-2l\ln\left[2\sin\left({(2l+1)\pi\over p}\right)\right]
-2l\ln\left[2\sin\left({(2l-1)\pi\over p}\right)\right]
\right]\cr
}\quad\; p=2r\no$$
\subsec{The Function $\tau_-(p,s)$ and $\ln\Det\dstar d_1$}
Let us now turn to the evaluation of $\tau_{-}(p,s)$.
$$\tau_{-}(p,s)=\sum_{n=1}^{\infty}
{n\left\langle\chi^{n+1\over2}\right\rangle_p
+(n+2)\left\langle\chi^{n-1\over2}\right\rangle_p
\over {(n+1)}^{2s}}
\no$$
\par
Decomposing the sum over $n$ into the different conjugacy classes
and using our general character formula we have
$$\eqalign{\tau_{-}(p,s)=&\sum_{k=1}^{\infty}{pk(pk+1)
+(pk+2)2k\over{(pk+1)}^{2s}}\cr
&\qquad+\sum_{k=1}^{\infty}{(pk-2)2k+pk(2k-1)\over{(pk-1)}^{2s}}\cr
&\qquad+\sum_{l=2}^{r-1}{2(pk-2l+1)(2k-1)\over{(pk-2l+1)}^{2s}}\cr
}\no$$
After some rearrangement we arrive at
$$\eqalign{\tau_{-}(p,s)=&\sum_{k=1}^{\infty}\left\{
{4\over p}\sum_{l=0}^{r-1}{1\over{(pk-2l+1)}^{2s-2}}\right.\cr
&\qquad+(1-{4\over p})
\left[{1\over{(pk+1)}^{2s-1}}-{1\over{(pk-1)}^{2s-1}}\right]\cr
&\qquad-\left[{1\over{(pk+1)}^{2s}}+{1\over{(pk-1)}^{2s}}\right]\cr
&\qquad+\left.{4\over p}
\sum_{l=2}^{r-1}(2l-1-r){1\over{(pk-2l+1)}^{2s-1}}\right\}\cr
}\no$$
Since the first term involves a sum over all odd conjugacy classes we
have
$$\eqalign{\sum_{k=1}^{\infty}\sum_{l=0}^{r-1}{1\over{(pk-2l+1)}^{2s-2}}
&=\sum_{m=1}^{\infty}{1\over{(2m+1)}^{2s-2}}\cr
&=(1-{1\over {2}^{2s-2}})\zeta(2s-2)-1\cr
}\no$$
Thus
\eqlabel{\tminp}
$$\eqalign{\tau_{-}(p,s)
=&{4\over p}\left((1-{1\over{2}^{2s-2}})\zeta(2s-2)-1\right)\cr
&\qquad+(p-4){1\over {p}^{2s}}
[\zeta(2s-1,1+{1\over p})-\zeta(2s-1,1-{1\over p})]\cr
&\qquad-{1\over {p}^{2s}}
[\zeta(2s,1+{1\over p})+\zeta(2s,1-{1\over p})]\cr
&\qquad-{2\over{p}^{2s}}
\sum_{l=2}^{r-1}\left(p-(2l-1)\right)\zeta(2s-1,1-{2l-1\over p})\cr
}\no$$
Noting that $p-(2l-1)$ ranges from $-(p-6)$ to $(p-6)$ in steps of
$2$ when $l$ ranges from $2$ to $r-1$
the final sum in \docref{tminp} is therefore of the form
$$\eqalign{\sum_{l=2}^{r-1}
\left(p-(2l-1)\right)\zeta(2s-1,1-{2l-1\over p})
&=\sum_{l=2}^{[{r-1\over
2}]}\left(p-2(2l-1)\right)\left[\zeta(2s-1,1-{2l-1\over
p})\right.\cr
&\left.-\zeta(2s-1,{2l-1\over p})\right]\cr
}\no$$
 Substituting back we obtain
\eqlabel{\tmns}
$$\eqalign{\tau_{-}(p,s)
&={4\over p}\left((1-{1\over{2}^{2s-2}})\zeta(2s-2)-1\right)\cr
&\qquad+(p-4){1\over {p}^{2s}}
[\zeta(2s-1,1+{1\over p})-\zeta(2s-1,1-{1\over p})]\cr
&\qquad-{1\over {p}^{2s}}
[\zeta(2s,1+{1\over p})+\zeta(2s,1-{1\over p})]\cr
&\qquad+{2\over{p}^{2s}}
\sum_{l=2}^{[{r-1\over 2}]}
\left(p-(2l-1)\right)
\left[\zeta(2s-1,{2l-1\over p})-\zeta(2s-1,1-{2l-1\over p})\right]\cr
}\no$$

Our first observation  is that
using $\zeta(-2)=0$,\quad
$\zeta(-1,1+a)-\zeta(-1,1-a)=-a$,\quad and
$\zeta(0,1+a)+\zeta(0,1-a)=-1$.\quad
$$\tau_{-}(p,0)=0\no$$
Now differentiation of \docref{tmns} with respect to $s$ and evaluating the
expression at $s=0$, and using some of our relations from Appendix C gives
$$\eqalign{\tm(p)
&=-{24\over p}\zeta '(-2)
+2(p-4)[\zeta '(-1,{1\over p})-\zeta '(-1,1-{1\over p})]
-2[\zeta '(0,{1\over p})+\zeta '(0,1-{1\over p})]\cr
&\qquad+4\sum_{l=1}^{[{r-3\over 2}]}
\left(p-(2l-1)\right)
\left[\zeta '(-1,{2l-1\over p})-\zeta '(-1,1-{2l-1\over p})\right]\cr
}\no$$
\par
Further use or our the relations derived in Appendix C  allows us
to express the result in terms of integrals over
trigonometric functions as in the preceding sections to yield
\eqlabel{\taumfin}
$$\eqalign{\tm(p)
&=4\ln(2\sin{\pi\over p})+{6\over p{\pi}^2}\zeta(3)
+2(p-4)
[{1\over\pi}\int_{0}^{\pi\over p}dz z
\cot(z)-{1\over p}\ln(2\sin{\pi\over p})]\cr
&\quad+4\sum_{l=1}^{[{r-3\over 2}]}
\left(p-(2l+1)\right)
\left[{1\over\pi}\int_{0}^{{\pi(2l+1)\over p}}dz z
\cot(z)-{(2l+1)\over p}\ln(2\sin{\pi(2l+1)\over p})\right]\cr
}\no$$
This again decomposes into the two cases $p=0,2\mod4$ and
these yield the expressions
$$\eqalign{\tm(p)
&=4\ln(2\sin{\pi\over p})+{6\over p{\pi}^2}\zeta(3)
+2(p-4)
[{1\over\pi}\int_{0}^{\pi\over p}dz z
\cot(z)-{1\over p}\ln(2\sin{\pi\over p})]\cr
&+4\sum_{l=1}^{t-1}
\left(p-(2l+1)\right)
\left[{1\over\pi}\int_{0}^{{\pi(2l+1)\over p}}dz z
\cot(z)-{(2l+1)\over p}\ln(2\sin{\pi(2l+1)\over p})\right]\cr
& \hskip0.8\hsize p=2\mod4\cr}\no$$
and
$$\eqalign{\tm(p)
&=4\ln(2\sin{\pi\over p})+{6\over p{\pi}^2}\zeta(3)
+2(p-4)
[{1\over\pi}\int_{0}^{\pi\over p}dz z
\cot(z)-{1\over p}\ln(2\sin{\pi\over p})]\cr
&+4\sum_{l=1}^{t-1}
\left(p-(2l+1)\right)
\left[{1\over\pi}\int_{0}^{{\pi(2l+1)\over p}}dz z
\cot(z)-{(2l+1)\over p}\ln(2\sin{\pi(2l+1)\over p})\right]\cr
& \hskip0.8\hsize p=0\mod4\cr}\no$$
\subsec{The Torsion $T(p)$ and $\tau(p,s)$.}
We can now combine our results for $\tau_{+}(p,0)$
and $\tau_{-}(p,0)$ to obtain expressions for the torsion in the present
case where $p$ is even. Note again that since $\tau_{\pm}(p,0)=0$ we have
$$\tau(p,0)=0\no$$
for the case of $p$ even, and again  this ensures that the torsion is
metric independent.
\par
Combining the expressions \docref{taupfin} and \docref{taumfin}
for $\tp(p)$ and $\tm(p)$ we obtain two expressions
for $\ln T(p)$: one for each conjugacy class; these are
$$\eqalign{\ln T(p)
&=-4\ln(2\sin{\pi\over p})+[{2(p-4)\over p}\ln(2\sin{\pi\over
p})]\cr
&\qquad+{4\over p}
\sum_{l=1}^{(p-6)/4}\left[{(2l+1)}^2-2\right]
\ln\left[2\sin\left({(2l+1)\pi\over
p}\right)\right]\cr
&\qquad
-{8\over\pi}\sum_{l=1}^{(p-6)/4}(2l+1)
\int_{0}^{(2l+1)\pi\over p}dz\,z\cot(z)\cr
}\qquad p=2\mod4\no$$
and
$$\eqalign{\ln T(p)
&=-4\ln(2\sin{\pi\over p})
-{4\over p}\ln2+2(p-4)[{1\over p}\ln(2\sin{\pi\over p})]\cr
&\qquad+{4\over p}\sum_{l=1}^{(p-4)/4}\left[(2l+1)(2l+1)-2\right]
\ln\left[2\sin\left({(2l+1)\pi\over p}\right)\right]\cr
&\qquad-{4\over\pi}\sum_{l=1}^{(p-4)/4}(2l-1)
\int_{0}^{(2l+1)\pi\over p}dz\,z\cot(z)\cr
}\qquad p=0\mod4\no$$
We note that for the above two formulae the torsion is already given by
the first term on their RHS's and so we obtain  somewhat non-trivial
integration formulae for the integrals therein. On using the relations
derived at the end of appendix B we can also reduce the above
expressions to Ray's expression for the torsion.
\par
We have therefore now obtained a complete list of the
determinants of Laplacians for 0 and 1 forms on the lens spaces
$L(p)$ for all integer $p\ge 2$, as well as the torsion $T(p)$ for all $p\ge2$.
\beginsection{Conclusion}
In the preceeding sections we analysed by direct computation the
the determinants of Laplacians on 0 and 1-forms on the lens spaces $L(p)$,
defined via the derivatives  of their associated  zeta functions.
\par
In this concluding section we collect our results and present in
graphical form the behaviour of the sequence of determinants we have
analysed  and their related torsion.
\par
For 0-forms we found that
$$\eqalign{\ln\Det d^*d_0=&{1\over 2p\pi^2}\zeta(3)
+\ln\left[2\sin({\pi\over p})\right]
+{(p-2)\over\pi}\int_{0}^{\pi/p}\ln\left[2\sin(z)\right]\cr
&\qquad-{2\over \pi}\sum_{l=1}^{(p-3)/2}(2l)\int_{0}^{2l\pi/p}
\ln\left[2\sin(z)\right]\cr
}\qquad p=3,5,\ldots\no$$
in the case of $p$ odd and
$$\eqalign{\ln\Det d^*d_{0}
&=-{3\over p{\pi}^2}\zeta(3)
+{(p-4)\over\pi}\int_{0}^{\pi\over p}dz\ln\left[2\sin(z)\right]\cr
&+2\sum_{l=1}^{t-1}
\left(p-(2l+1)\right)
{1\over\pi}\int_{0}^{{\pi(2l+1)\over p}}dz\,
\ln\left[2\sin(z)\right]\cr
}\qquad p=2\mod4\no$$
 and
$$\eqalign{\ln\Det d^*d_{0}
&=-{6\over p{\pi}^2}\zeta(3)\cr
&\qquad+
{(p-4)\over\pi}\int_{0}^{\pi\over p}dz\,\ln\left[2\sin(z)\right]\cr
&\qquad+2\sum_{l=1}^{t-1}
\left(p-(2l+1)\right){1\over\pi}\int_{0}^{{\pi(2l+1)\over p}}dz\,
\ln\left[2\sin(z)\right]\cr
}\qquad p=0\mod4\no$$
\par
These are plotted in {\it figure 1}.
\par
While for 1-forms we found
$$\eqalign{\ln\Det d^*d_1=&{1\over p\pi^2}\zeta(3)
-2\ln\left[2\sin({\pi\over p})\right]
+{2(p-2)\over\pi}\int_{0}^{\pi/p}\ln\left[2\sin(z)\right]\cr
&\qquad-{4\over \pi}\sum_{l=1}^{(p-3)/2}(2l)\int_{0}^{2l\pi/p}
\ln\left[2\sin(z)\right]\cr
}\qquad p=3,5,\ldots\no$$
while when $p$ is even we found that
$$\eqalign{\ln\Det d^*d_{1}
&=-4\ln(2\sin{\pi\over p})-{6\over p{\pi}^2}\zeta(3)
+2(p-4)
{1\over\pi}\int_{0}^{\pi\over p}dz\,\ln\left[2\sin(z)\right]\cr
&+4\sum_{l=1}^{t-1}
\left(p-(2l+1)\right)
{1\over\pi}\int_{0}^{{\pi(2l+1)\over p}}dz\,
\ln\left[2\sin(z)\right]\cr
}\qquad p=2\mod4\no$$
 and
$$\eqalign{\ln\Det d^*d_{1}
&=-4\ln(2\sin{\pi\over p})-{6\over p{\pi}^2}\zeta(3)\cr
&\qquad+2(p-4)
{1\over\pi}\int_{0}^{\pi\over p}dz\,
\ln\left[2\sin(z)\right]\cr
&\qquad+4\sum_{l=1}^{t-1}
\left(p-(2l+1)\right)
{1\over\pi}\int_{0}^{{\pi(2l+1)\over p}}dz\,
\ln\left[2\sin(z)\right]\cr
}\qquad p=0\mod4\no$$
 and these results are displayed in {\it figure  2}.
The {\it difference}
$$\eqalign{\ln T(p)&=\ln\Det d^*d_1-2\ln\Det d^*d_0\cr
&=-4\ln\left[2\sin({\pi\over p})\right]\cr
}\no$$
which gives the torsion itself,
is plotted in {\it figure 3}.
\par
A perusal of figure 3 shows that $\ln T(p)$ is negative for small $p$
and large and positive for large $p$. This raises the question as to
whether $\ln T(p)$ crosses the $p$ axis at an integer value or not. If
so this corresponds to a trivial value for the torsion. In fact this
clearly does happen for the value $p=6$ i.e. we have
$$\ln T(6)=0\no$$
We show the more detailed behaviour of the torsion for small $p$ in
{\it figure 4}.
\par
Further interesting
results are that that if we
work with $L(p,q)$
rather than $L(p)\equiv L(p,1)$ then  the torsion, now denoted by $T(p,q)$, is
trivial for {\it only} two other
three dimensional lens spaces, namely $L(10,3)$ and $L(12,5)$: we find
that
$$\eqalign{\ln T(p,q)&=-2 \ln\left[4\sin\left({\pi\over p}\right)
\sin\left({\pi q^*\over p}\right)\right]\cr
\hbox{where }q^*& \hbox{ satisfies }q q^*=1\,\mod p\cr}\no$$
It is then possible to prove that, for $p>12$, $\ln T(p,q)$ is {\it
strictly positive}; while for $p\le12$ a check of the finite number of
cases yields triviality in just the three cases given above.
We conjecture that  this may be understandable using some form  of
supersymmetry. These formulae have yet to be elucidated further.
\par
The precise meaning of our formulae such as
$$\eqalign{\zeta(3)&={2\pi^2\over7}\ln(2)
-{8\over7}\int_0^{\pi/2}dz\,z^2\cot(z)\cr
\zeta(3)&={2\pi^2\over13}\ln 3
-{9\over 13}\int_{0}^{\pi\over3}dz\,z(z+{\pi\over3})\cot(z)
-{9\over 13}\int_{0}^{2\pi\over3}dz\,z(z-{\pi\over3})\cot(z)
\cr}\no$$
and the more general
$$\eqalign{\zeta(3)&={2\pi^2\over (p^3-1)}\left[
2(p-1)\ln\left[2\sin({\pi\over p})\right]
-2\sum_{l=1}^{(p-3)\over 2}
4l^2\ln\left[2\sin({2l\pi\over p})\right] \right.\cr
&\qquad-{p^2\over\pi^2}\int_{0}^{\pi\over p}dz\,
z(z+{(p-2)\pi\over p})\cot(z)
-{p^2\over\pi^2}\int_{0}^{(p-1)\pi\over p}dz\,
z(z-{(p-2)\pi\over p})\cot(z)\cr
&\qquad\left.-2\sum_{l=1}^{(p-3)\over 2}
{p^2\over\pi^2}\int_{0}^{2l\pi\over p}dz\, z(z-{4l\pi\over p})\cot(z)
\right],\qquad p=3,5,\ldots\cr
}\no$$
is, as yet, unclear. There may be some number theoretic matters
underlying them as seems to be the case in other work on lens spaces,
cf. \ref{15}.
A thought provoking fact is
that $\zeta(3)$ occurs in a recent paper of Witten \ref{16} where, after
multiplication by a known constant, it gives the {\it volume} of the
symplectic space of flat connections over a {\it non-orientable} Riemann
surface. The corresponding calculation for  {\it orientable} surfaces
(where the volume element is a rational cohomology class)
allows a cohomological rederivation of the irrationality of $\zeta(2),\,
\zeta(4),\dots$. This paper also involves the torsion but in two dimensions
rather than three.
The proof that $\zeta(3)$ is irrational was only obtained in 1978 cf. \ref{17}
and the rationality of $\zeta(5),\,\zeta(7),\ldots$ is at present open.
However there are now other proofs \ref{18}, one of which uses the
characters of conformal quantum field theory.
\par
Finally, our technique, applied in five dimensions instead of three, would
yield
formulae for $\zeta(5)$ but their nature has not yet been explored.
\par\vfill\eject


\vsize=1.5\vsize
\centerline{\bf FIGURES: The Determinants}
\par\vskip6truein
\special{dvitops: import tpluseve.ps.keep 6in \the\hsize}
\par\vskip-1.75truein
\centerline{{\bf Figure 1:} $2\ln \dstar d_0$ versus $p$.}
\par\vskip6truein
\special{dvitops: import tplusodd.ps.keep 6in \the\hsize}
\par\vskip-1.75truein
\centerline{{\bf Figure 2:} $\ln \dstar d_1$ versus $p$.}
\vsize=.666\vsize
\par\vfill\eject
\vsize=1.5\vsize
\centerline{\bf FIGURES: The Torsion}
\par\vskip6truein
\special{dvitops: import rtor100.ps.keep 6in \the\hsize}
\par\vskip-1.75truein
\centerline{{\bf Figure 3:} $\ln T(p)$ versus $p$.}
\par\vskip6truein
\special{dvitops: import smallptorsion.ps.keep 6in \the\hsize}
\par\vskip-1.75truein
\centerline{{\bf Figure 4:} $\ln T(p)$ versus $p$ for small p.}
\vsize=.666\vsize
\par\vfill\eject
\beginappendix{The function $A_{p}(x,s)$}
In this appendix we analyze the function $$A_{p}(x,s)
=\sum_{k=1}^{\infty}{{(pk+x)}^2\over{\left\{{(pk+x)}^2-1\right\}}^s}\no$$
We are interested in particular in the value of this function and
its derivative with respect to $s$ at $s=0$. For this purpose we
denote
$$A_{p}(k,x,s)={{(pk+x)}^2\over{\left\{{(pk+x)}^2-1\right\}}^s}\no$$
which has the expansion $$\eqalign{A_{p}(k,x,s) &={1\over
{(pk)}^{2s-2}}\left[1+2(1-s){x\over pk}
+\left(s+(s-1)(2s-1)x^2\right){1\over {(pk)}^2}\right.\cr
&\qquad\left.
-2sx\left(s+{(s-1)(2s-1)\over3}x^2\right){1\over{(pk)}^3}+\dots\right]
\cr }\no$$ Summing over $k$ leads to $$\eqalign{A_{p}(x,s)
&={1\over{p}^{2s-2}}\zeta(2s-2)+{2x\over{p}^{2s-1}}(1-s)\zeta(2s-1)
+{1\over{p}^{2s}}\left(s+(1-s)(1-2s)x^2\right)\zeta(2s)\cr
&\qquad-{1\over{p}^{2s+1}}\left(s+{(1-s)(1-2s)\over3}x^2\right)x2s\zeta(2s+1)
+\hat A_{p}(x,s)\cr }\no$$
where
$$\eqalign{\hat A_{p}(x,s)
&=\sum_{k=1}^{\infty}\left\{
{{(pk+x)}^2\over{\left\{{(pk+x)}^2-1\right\}}^s} -{1\over
{(pk)}^{2s-2}}\left[1+2(1-s){x\over pk}\right.\right.\cr
&\left.\left.+\left(s+(s-1)(2s-1)x^2\right){1\over {(pk)}^2}
-2sx\left(s+{(s-1)(2s-1)\over3}x^2\right){1\over{(pk)}^3}
\right]\right\}\cr }\no$$ Now the function $\hat A_{p}(x,s)$ is such
that the processes of summation and differentiation with respect to
$s$ at $s=0$ commute. Also it is such that $\hat A_{p}(x,0)=0$,
which yields $$A_{p}(x,0)=-{x(x+p)(2x+p)\over 6p}\no$$

Next evaluating the derivative at $s=0$ we have
$$\eqalign{A_{p}'(x,0) &=-A_{p}(x,0)\ln{p}^2+2{p}^{2}\zeta
'(-2)+\zeta(0) +2px\left[2\zeta '(-1)-\zeta(-1)\right]\cr
&\qquad+\left[2\zeta '(0)-3\zeta(0)\right]x^2 -{x\over
p}+{1\over{3p}}\left[3-2\gamma\right]x^3 +\hat A_{p}'(x,0)\cr }\no$$
which on using $\zeta(0)=-{1\over 2}$, $\zeta(-1)=-{1\over12}$,
$2s\zeta(2s+1)=1+2\gamma s+\dots$ and
$\zeta'(0)=-{1\over2}\ln{2\pi}$ we have
$$\eqalign{A_{p}'(x,0)&=-A_{p}(x,0)\ln{p}^2+2p^2\zeta
'(-2)-{1\over2} +\left[4p\zeta'(-1)+{p\over6}-{1\over p}\right]x\cr
&\qquad+\left[{3\over2}-\ln{[2\pi]}\right]x^2 +{1\over
3p}\left[3-2\gamma\right]x^3+\hat A_{p}'(x,0)\cr }\no$$ Since the
processes of differentiation with respect to $s$ and performing the
sum over $k$ commute for $\hat A_{p}(x,s)$ we analyze this function
by first taking the derivative and then performing the sum.
$$\eqalign{\hat A_{p}'(k,x,0) &=-{(pk
+x)}^2\left[\ln\left[1+{(x+1)\over pk}\right]
+\ln\left[1+{(x-1)\over pk}\right]\right]\cr
&+{(pk)}^2\left[{2x\over pk}-{(1-3x^2)\over
{(pk)^2}}+{2x^3\over{3(pk)}^3}\right]\cr }\no$$ We note that
$$\eqalign{\ln{(1+{(x+1)\over pk})}+\ln{(1+{(x-1)\over pk})}
&=\sum_{m=1}^{\infty} {{(-1)}^{(m+1)}\,\left[{(x+1)}^{m}
+{(x-1)}^{m}\right]\over m\,\,{(pk)}^{m}}\cr }\no$$ Hence combining
this with  $\hat A_{p}'(k,x,0)$  gives $$\eqalign{\hat A_{p}'(k,x,0)
&=\sum_{m=4}^{\infty}
{{(-1)}^{m}\left[{(1+x)}^{m}+{(x-1)}^{m}\right] \over
m\,\,{(pk)}^{(m-2)}}\cr &\qquad+2x\sum_{m=3}^{\infty}
{{(-1)}^{m}\left[{(1+x)}^{m}+{(x-1)}^{m}\right] \over
m\,\,{(pk)}^{(m-1)}}\cr &+x^2\sum_{m=2}^{\infty}
{{(-1)}^{m}\left[{(1+x)}^{m}+{(x-1)}^{m}\right] \over
m\,\,{(pk)}^{m}}\cr }\no$$ which gives $$\eqalign{\hat A_{p}'(x,0)
&=\sum_{m=2}^{\infty}
{{(-1)}^{m}\left[{(1+x)}^{(m+2)}+{(x-1)}^{(m+2)}\right] \over
(m+2)\,\,{p}^{m}}\zeta(m)\cr &\qquad-2x\sum_{m=2}^{\infty}
{{(-1)}^{m}\left[{(1+x)}^{(m+1)}+{(x-1)}^{(m+1)}\right] \over
(m+1)\,\,{p}^{m}}\zeta(m)\cr &+x^2\sum_{m=2}^{\infty}
{{(-1)}^{m}\left[{(1+x)}^{m}+{(x-1)}^{m}\right] \over
m\,\,{p}^{m}}\zeta(m)\cr }\no$$

We now observe the identity
$$\sum_{m=2}^{\infty}{(-1)}^{m}{a}^{m-1}\zeta(m,\alpha)
=\psi(\alpha+a)-\psi(\alpha)
\no$$
which  yields
\eqlabel{\sumform}
$$
\eqalign{\sum_{m=2}^{\infty}{{(-1)}^{m}{a}^{m+\nu}\over
m+\nu}\zeta(a,\alpha)
&=\int_{0}^{a}dy\,y^{\nu}\left[\psi(y+\alpha)-\psi(\alpha)\right]\cr
}\no$$
for $\nu\geq 0$.
Now using this  identity we have
$$\eqalign{\hat A_{p}'(x,0)&=p^2\int_{0}^{(x+1)/ p}dy\,
{(y-{x\over p})}^2
\left[\psi(1+y)-\psi(1)\right]\cr
&\qquad+p^2\int_{0}^{(x-1)/ p}dy\,
{(y-{x\over p})}^2\left[\psi(1+y)-\psi(1)\right]\cr
}\no$$

Finally using $\psi(1)=-\gamma$
we arrive at at our expression for the desired
analytic continuation of the derivative at zero,
\eqlabel{\Aprimex}
$$\eqalign{A_{p}'(x,0)=&-A_{p}(x,0)\ln{p}^2 -{p^2\over
2\pi^2}\zeta(3)-{1\over2} +{p x\over6}
\left[24\zeta '(-1)+1-{6\over p^2}\right]
+x^2({3\over 2}- \ln[2\pi])\cr
&\qquad+{x^3\over p}
+p^2\int_{0}^{(x+1)\over p}{(y-{x\over p})}^2\psi(1+y)
+p^2\int_{0}^{(x-1)\over p}{(y-{x\over p})}^2\psi(1+y)\cr}\no$$
We are interested in particular in the symmetric sum
$$A_{p}^{+}(x)=A_{p}'(x,0)+A_{p}'(-x,0)\no$$
which from \docref{Aprimex}  on using
\eqlabel{\gamreln}
$$\Gamma(1+x)\Gamma(1-x)={\pi x\over\sin(\pi x)}\no$$
we find to be
\eqlabel{\ApluspA}
$$\eqalign{A_{p}^{+}(x)
=-{p^2\over\pi^2}\zeta(3)
&-{p^2\over\pi^2}\int_{0}^{(x+1)\pi\over p}
dz\, z(z-{2\pi x\over p})\cot(z)
-{p^2\over\pi^2}\int_{0}^{(x-1)\pi\over p}
dz\, z(z-{2\pi x\over p})\cot(z)\cr
&-x^2\ln\left[{2\sin\left({(x+1)\pi\over p}\right)\over x+1}\right]
-x^2\ln\left[{2\sin\left({(x-1)\pi\over p}\right)\over x-1}\right]\cr
}\no$$
Alternatively one can see this  directly from the expansion of
$\hat A_{p}^{+}$ in the series representation and the identities
$$\sum_{l=1}^{\infty}{1\over l+1}{1\over a^{2l}}\zeta(2l)
={1\over2}-{a^2\over{\pi}^2}\int_{0}^{\pi\over a}dz z^2\cot(z)
\no$$
$$\sum_{l=1}^{\infty}{1\over 2l+1}{1\over{a}^{2l}}\zeta(2l)
={1\over2}-{a\over 2\pi}\int_{0}^{\pi\over a}dz z\cot(z)
\no$$
$$\sum_{l=1}^{\infty}{1\over l}{1\over{a}^{2l}}\zeta(2l)
=-\ln\left[{\sin({\pi\over a})\over{\pi\over a}}\right]
\no$$

A useful expression is obtained by summing over all conjugacy
classes to obtain an expression independent of $p$.
Hence note that
$$\eqalign{\sum_{j=0}^{p-1}\sum_{k=1}^{\infty}A_{p}(k,-j,s)
&=\sum_{\{\nu_{p}\}}\sum_{k=1}^{\infty}
{{(pk-j)}^2\over{\{{(pk-j)}^2-1\}}^s}\cr
&=\sum_{n=1}^{\infty}{{(n+1)}^2\over{\{{(n+1)}^2-1\}}^s}\cr }\no$$
is independent of $p$ since we have merely decomposed it as a sum
over conjugacy classes mod $p$. We label this sum $A(s)$ and its
derivative at $s=0$ simply $A$.

Now denoting
$$\eqalign{A_{p}
&=A_{p}'(0,0)
+\sum_{l=0}^{(p-3)\over2}A_{p}^{+}(2l+1)\cr}$$
As a consequence of the above we have that
$$A=A_{p}-\sum_{l=1}^{(p-3)\over2}{(2l+1)}^2\ln[{(2l+1)}^2-1]\no$$
is independent of $p$. This invariant can be written as
$$\eqalign{ A&=-{p^3\over2\pi^2}\zeta(3)
-\sum_{l=1}^{(p-3)\over2}{(2l+1)}^2\ln[{(2l+1)}^2-1]
-{p^2\over\pi^2}\int_{0}^{\pi\over p}dz z^2 \cot(z)\cr
&-{p^2\over\pi^2}\sum_{l=0}^{(p-3)\over2}
\left[\int_{0}^{(2l+2)\pi\over p}dz z(z-{2\pi(2l+1)\over p})\cot(z)
+\int_{0}^{2l\pi\over p}dz z(z-{2\pi(2l+1)\over p})\cot(z)\right]\cr
&-\sum_{l=0}^{(p-3)\over2}{(2l+1)}^2
\left[\ln\left[{2\sin\left({(2l+2)\pi\over p}\right)\over 2l+2}\right]
+\ln\left[{2\sin\left({2l\pi\over p}\right)\over 2l}\right]\right]\cr
}\no$$
Which on simplifying gives
$$\eqalign{A
&=-{p^3\over 2\pi^2}\zeta(3)
-{(p-2)}^2\ln[2\sin({\pi\over p})]-\ln[{\pi\over p}]
-{p^2\over\pi^2}\int_{0}^{\pi\over p}dz z^2\cot(z)\cr
&\qquad-{p^2\over\pi^2}\int_{0}^{(p-1)\pi\over p}dz
z(z-{2(p-2)\pi\over p})\cot(z)\cr
&\qquad-2\sum_{l=1}^{(p-3)\over2}\left[
{p^2\over\pi^2}\int_{0}^{2l\pi\over p}dz z(z-{4l\pi\over p})\cot(z)
 +(4l^2+1)\ln[2\sin({2l\pi\over p})]\right]\cr
}\no$$
This constant has the value $A=-1.20563$

One can use this together with the expression for $B_{p}^{-}(x)$
obtained in the next appendix to  give an alternative form of the expression
for $\tp(p)$. Explicitly
$$\eqalign{\tp(p)=&{2\over p}A_{p}'(0,0)
+{2\over p}\sum_{l=0}^{(p-3)/2}
\left[A_{p}^{+}(2l+1)-(2l+1)B_{p}^{-}(2l+1)\right]\cr
=&{2\over p}\left[A
-\sum_{l=1}^{(p-3)/p}(2l+1)\left[B_{p}(2l+1)-(2l+1)
\ln\left[{(2l+1)}^2-1\right]\right]\right]}$$
which works out to be
$$\eqalign{\tp(p)
&={2\over p}\left[A
+\ln [{\pi\over p}]
+{(p-2)}^2\ln[2\sin({\pi\over p})]-{(p-2)p\over
\pi}\int_{0}^{(p-1)\pi\over p}dz z\cot(z) \right.\cr
&\qquad\left.-2{p\over
\pi}\sum_{l=1}^{(p-3)\over2}2l\int_{0}^{2l\pi\over p}dz z
\cot(z)
+\sum_{l=1}^{(p-3)\over2}(4l^2+1)
\ln\left[2\sin({2l\pi\over p})\right]\right]\cr
}\no$$

\beginappendix{The function $B_{p}(x,s)$}
In this appendix we analyze the function
$$B_{p}(x,s)
=\sum_{k=1}^{\infty}{(pk+x)\over{\left\{{(pk+x)}^2-1\right\}}^s}\no$$
We are again interested in the value of this function and its
derivative with respect to $s$ at $s=0$. For this we  denote
$$B_{p}(k,x,s)={(pk+x)\over{\left\{{(pk+x)}^2-1\right\}}^s}\no$$
The continuation now requires us to
extract the appropriate large $k$ behaviour from $B_{p}(x,s)$.
Proceeding to do this we find
$$\eqalign{B_{p}(k,x,s)=&
{(pk)}^{1-2s}-(2s-1)x{(pk)}^{-2s}+s(1+(2s-1)x^2){(pk)}^{-2s-1}
+\hat B_{p}(k,x,s)\cr
}\no$$
Hence summing over $k$ yields
$$B_{p}(x,s)={1\over {p}^{2s}}\left[p\zeta(2s-1)
-x(2s-1)\zeta(2s)+{s\left(1+(2s-1)x^2\right)\over
p}\zeta(2s+1)\right]+\hat B_{p}(s,x)\no$$
where
$$\eqalign{\hat B_{p}(x,s)&=\sum_{k=1}^{\infty}\left[
{{(pk+x)}\over{\left\{{(pk+x)}^2-1\right\}}^s}
-{1\over {(pk)}^{2s-1}}\left\{1-x{(2s-1)\over pk}\right.\right.\cr
&\left.\left.+{s\left(1+(2s-1)x^2\right)\over{(pk)}^2}
-{x\over3}{s(2s+1)\left(3+(2s-1)x^2\right)\over {(pk)}^3}
\right\}\right]\cr
}\no$$
$\hat B_{p}(x,s)$ has the property that the process of taking the
limit differentiating with respect to $s$ and taking $s\rightarrow
0$ commutes with the sum over $k$. The second property is that
$\hat B_{p}(x,0)=0$.
Hence this expression has the analytic continuation to $s=0$
$$\eqalign{B_{p}(x,0)&=p\zeta(-1,1+{x\over p})+{1\over 2p}\cr
&=-{p\over 12}-{x\over 2}+{(1-x^2)\over 2p}\cr
}\no$$
Differentiating with respect to $s$ and evaluating at $s=0$ yields
$$\eqalign{B_{p}'(x,0)&=
-B_{p}(x,0)\ln p^2 +2p\zeta '(-1)+2x\left[\zeta '(0)-\zeta(0)\right]
+{x^2\over p}+{\gamma(1-x^2)\over p}
+ \hat B_{p}'(x,0)\cr
}\no$$
We choose to do the evaluate $\hat B_{p}'(x,0)$ by doing the
derivative first and the sum last.  Therefore
$$\eqalign{\hat B_{p}'(k,x,0)&=\sum_{m=2}^{\infty}
{{(-1)}^{(m+1)}\left[{(1+x)}^{(m+1)}+{(x-1)}^{(m+1)}\right]
\over (m+1)\,\,{pk}^{m}}\cr
&\qquad+x\sum_{m=2}^{\infty}
{{(-1)}^{m}\left[{(1+x)}^{m}+{(x-1)}^{m}\right]
\over m\,\,{pk}^{m}}\cr
}\no$$
and
$$\eqalign{\hat B_{p}'(x,0)&=\sum_{m=2}^{\infty}
{{(-1)}^{(m+1)}\left[{(1+x)}^{(m+1)}+{(x-1)}^{(m+1)}\right]
\over (m+1)\,\,{p}^{m}}\zeta(m)\cr
&\qquad+x\sum_{m=2}^{\infty}
{{(-1)}^{m}\left[{(1+x)}^{m}+{(x-1)}^{m}\right]
\over m\,\,{p}^{m}}\zeta(m)\cr
}\no$$
Again using the summation \docref{sumform} as in appendix A we have
$$\eqalign{\hat B_{p}'(x,0)&=-p\,\int_{0}^{(x+1)/ p}dy\,
(y-{x\over p})\left(\psi(1+y)-\psi(1)\right)\cr
&\qquad-p\,\int_{0}^{(x-1)/ p}dy\,
(y-{x\over p})\left(\psi(1+y)-\psi(1)\right)\cr}
\no$$
Thus we arrive at
$$\eqalign{B_{p}'(x,0)&=-B_{p}(x,0)\ln p^2
+2p\zeta '(-1)+\left[1-\ln[2\pi]\right]x
+{x^2\over p}\cr
&-p\,\int_{0}^{(x+1)/ p}dy\,
(y-{x\over p})\psi(1+y)
-p\,\int_{0}^{(x-1)/ p}dy\,
(y-{x\over p})\psi(1+y)\cr
}\no$$

For the function $B_{p}$ it is the anti-symmetric part that
contributes to the quantities of interest thus defining
$$\eqalign{B_{p}^{-}(x)&=B_{p}'(x,0)-B_{p}'(-x,0)\cr
&=x\ln{p}^2+\left[1-\ln[2\pi]\right]2x+\hat B_{p}^{-}(x)\cr
}\no$$
we have
\eqlabel{\BminuspB}
$$\eqalign{B_{p}^{-}(x)
&= {p\over\pi}\int_{0}^{(x+1)\pi/p}dz\,z\cot(z)
+{p\over\pi}\int_{0}^{(x-1)\pi/ p}dz\, z\cot(z)\cr
&\qquad-x\ln\left[{2\sin\left({(x+1)\pi\over p}\right)\over x+1}\right]
-x\ln\left[{2\sin\left({(x-1)\pi\over p}\right)\over x-1}\right]\cr
}\no$$
\subsec{The function $H_{p}(x,s)$}
One can combine the results above with those in Appendix A to
obtain an expression for $H_{p}'(x,0)$, where we have defined
$$H_{p}(x,s)=A_{p}(x,s)-xB_{p}(x,s)\no$$
Explicitly we have
$$\eqalign{H_{p}'(x,0)&=-H_{p}(x,0)\ln{p}^2 -{p^2\over
2\pi^2}\zeta(3)-{1\over2} + {p x\over 6}\left(12\zeta '(-1) + 1 -
{6\over p^2}\right)\cr &\qquad+ {x^2\over2}+p^2\int_{0}^{(x+1)\over
p} y(y-{x\over p})\psi(1+y) +p^2\int_{0}^{(x-1)\over p}dy y
(y-{x\over p})\psi(1+y)\cr }\no$$
where
$$H_{p}(x,0)=-{px\over12}-{x\over 2p}+{x^3\over 6p}\no$$
By analogy with $A_{p}^{+}(x)$ we can define
$$H_{p}^{+}(x)=A_{p}^{+}(x)-xB_{p}^{-}(x)\no$$
which is given by
$$H_{p}^{+}(x)=-{p^2\over\pi^2}\zeta(3)
-{p^2\over\pi^2}\int_{0}^{(x+1)\pi/p}dz\,z(z-{\pi x\over p})\cot(z)
-{p^2\over\pi^2}\int_{0}^{(x-1)\pi/ p}dz\,z(z-{\pi x\over p})\cot(z)\no$$
\beginappendix{Expressions involving the Hurwitz zeta function.}
The task of this appendix is to obtain an expressions for
$\zeta '(0,1+a)$, $\zeta '(-1,1+a)$ and $\zeta '(-2,1+a)$. We begin with
$\zeta '(0,1+a)$.
Note first of all  that
$$\zeta(s,1+a)=\sum_{n=1}^{\infty}{1\over{(n+a)}^{s}}\no$$
has the series expansion
$$\zeta(s,1+a)=\zeta(s)+s\zeta(s+1)a+\hat\zeta(s,1+a)\no$$
where
$$\eqalign{\hat\zeta(s,1+a)&=\sum_{n=1}^{\infty}\left[{1\over{(n+a)}^s}
-{1\over n^s}+{sa\over n^{s+1}}\right]\cr
&=\sum_{k=2}^{\infty}{{(-1)}^{k}\Gamma(s+k)\over\Gamma(s)k!}
\zeta(s+k)a^k\cr
}\no$$
Which gives the analytic continuation for $\zeta(s,1+a)$ to $s=0$.
We therefore deduce
$$\zeta(0,1+a)=\zeta(0)-a\no$$
and
$$\zeta '(0,1+a)=-{1\over 2}\ln[2\pi]-\gamma a+\hat\zeta '(0,1+a)$$
where we used $\zeta '(0)=-{1\over2}\ln[2\pi]$ and
$s\zeta(s+1)=1+\gamma s+\dots$.
Differentiation with respect to $s$ at $s=0$ and summation over
$n$ commute for $\hat\zeta(s,1+a)$ we therefore have
$$\eqalign{\hat\zeta '(0,1+a)&=-\sum_{n=1}^{\infty}\left[
\ln[1+{a\over n}]-{a\over n}\right]\cr
&=\sum_{k=2}^{\infty}{{(-1)}^k a^k\over k}\zeta(k)\cr
}\no$$
Thus using \docref{sumform} we have
$$\eqalign{\hat\zeta '(0,1+a)
&=\int_{0}^{a}dy\,\left[\psi(1+y)-\psi(1)\right]\cr
}\no$$
We therefore obtain that
$$\zeta '(0,1+a)=-{1\over2}\ln[2\pi]+\ln\Gamma(1+a)\no$$
The symmetric part of this function on using \docref{gamreln} can
be expressed as is expressed as
\eqlabel{\zid}
$$\zeta '(0,1+a)+\zeta '(0,1-a)=-\ln\left[{2\sin(\pi a)\over
a}\right]\no$$
Finally by noting that
$$\zeta(s)=p^{-s}\sum_{j=0}^{p-1}\zeta(s,1-{j\over p})$$
is a decomposition over conjugacy classes of $\zeta(s)$
Differentiating and evaluating at $s=0$ we obtain
\eqlabel{\conjdec}
$$\zeta '(0)=-\sum_{j=0}^{p-1}\zeta(0,1-{j\over p})\ln p
+\sum_{j=0}^{p-1}\zeta '(0,1-{j\over p})$$
Now the first term is just a decomposition of $\zeta(0)$ over
conjugacy classes therefore \docref{conjdec} gives the identity
\eqlabel{\zdecom}
$$\sum_{j=0}^{p-1}\zeta '(0,1-{j\over p})=\zeta '(0)+\zeta(0)\ln p$$
which for $p$ odd gives
$$\sum_{l=1}^{(p-1)/2}
\ln\left[2\sin({2l\pi\over p})\right]={1\over2}\ln p$$
or
$$\sum_{l=0}^{(p-1)/2}
\ln\left[2\sin\left({2l+1)\pi\over p}\right)\right]={1\over2}\ln p$$
Now for $p$ even  \docref{zdecom} takes the form
\eqlabel{\zevn}
$$\sum_{l=0}^{r-1}\left[\zeta '(0,1-{l\over r})
+\zeta '(0,1-{2l+1\over 2r})\right]=\zeta '(0)+\zeta(0)\ln[2r]$$
The first term in \docref{zevn} is of the form of the original
expresson \docref{zdecom} and therefore the resulting expression is
$$\sum_{l=0}^{r-1}\zeta '(0,1-{2l+1\over 2r})=-{1\over2}\ln2$$
For $p=4t+2$ we  divide the $l$-summation into the two
ranges, $0$ to $t-1$, and, $t$
to $2t-1$;  rearanging the latter sum and using \docref{zid} then makes
then gives us
$$-\sum_{l=0}^{t-1}\ln\left[2\sin\left({2l+1)\pi\over
p}\right)\right]+\zeta '(0,{1\over2})=-{1\over2}\ln2$$
Now noting that $\zeta '(0,{1\over2})=-{1\over2}\ln2$ we have the
identity.
$$\sum_{l=1}^{(p-6)/4}\ln\left[2\sin\left({(2l+1)\pi\over
p}\right)\right] =0$$
We find for $p=4t$, by a similar procedure, that
$$\sum_{l=0}^{(p-4)/2}\ln\left[2\sin\left({(2l+1)\pi\over p}\right)\right]
={1\over 2}\ln2$$
These are  the first set of identities  we obtain by decomposing over conjugacy
classes.
\subsec{Relations involving $\zeta '(-1,1+a)$}
Next we turn to $\zeta '(-1)$, thus note that
$$\zeta(s-1,1+a)=\sum_{n=1}^{\infty}{1\over{(n+a)}^{(s-1)}}\no$$
has the series expansion in terms of $a$
$$\zeta(s-1,1+a)
=\sum_{k=0}^{\infty}{{(-)}^{k}\Gamma(s+k-1)\,a^k
\over\Gamma(s-1)k!}\zeta(k+s-1)\no$$
which on extracting the divergent part gives
$$\zeta(s-1,1+a)=\zeta(s-1)-(s-1)a\zeta(s)
+{s(s-1)a^2\over2}\zeta(s+1)+\hat\zeta(s-1,1+a)\no$$
where
$$\hat\zeta(s-1,1+a)
=\sum_{n=1}^{\infty}\left[{1\over{(n+a)^{s-1}}}-{1\over n^{s-1}}
+{(s-1)a\over n^s}-{s(s-1)a^2\over n^{s+1}}\right]\no$$
This is now expressed in terms of our prescription for analytic
continuation.
Hence we have
$$\zeta(-1,1+a)=-{1\over 12}-{a(1-a)\over 2}\no$$
where $\zeta(-1)=-{1/2}$, $\zeta(0)=-{1/2}$ and
$s\zeta(s+1)=1+\gamma s+\dots$ have been used.
Differentiation with respect to $s$ and setting it to zero
yields
$$\zeta '(-1,1+a)=\zeta '(-1)
+\left[1-\ln[2\pi]\right]{a\over2}+{(\gamma-1)a^2\over 2}
+\hat\zeta '(-1,1+a)\no$$
Since differentiation with respect to $s$ at $s=0$ and summation
over $n$ commute we have
$$\eqalign{\hat\zeta '(-1,1+a)
&=-(n+a)\ln(1+{a\over n})+a+{a^2\over 2n}\cr
&=-\sum_{k=2}^{\infty}{(-1)^{k}a^{k+1}\over k+1}\zeta(k)
+a\sum_{k=2}^{\infty}{{(-)}^{k}a^k\over k}\zeta(k)\cr
}\no$$
Now using the identity
\eqlabel{\sumnf}
$$\sum_{k=2}^{\infty}{{(-)}^{k}a^{k+\nu}\over k+\nu}
\zeta(k)=\int_{0}^{a}dy\,y^{\nu}\left(\psi(1+y)-\psi(1)\right)
\no$$
valid for $a<1$ to remain
away from the poles of $\psi(y)$ which occur at $y=-n$ for
$n=0,1,2,\dots$ we have
$$\hat\zeta '(-1,1+a)
=-\int_{0}^{a}dy\,(y-a)\left[\psi(1+y)-\psi(1)\right]
\no$$
and therefore
$$\zeta '(-1,1+a)=\zeta '(-1)+\left[1-\ln(2\pi)\right]{a\over2}
-{a^2\over 2}
-\int_{0}^{a}dy\,(y-a)\psi(1+y)\no$$
Finally we conclude that the anti-symmetrised sum is given by
$$\zeta '(-1,1+a)-\zeta '(-1,1-a)=
\left[1-\ln[2\pi]\right]a-\int_{0}^{a}dy\,(y-a)\left[\psi(1+y)-\psi(1-y)\right]\no$$
Now using
\eqlabel{\psirln}
$$\psi(1+y)-\psi(1-y)
=-{d\over dy}\ln\left[{\sin(\pi y)\over\pi y}\right]\no$$
we obtain
$$\zeta '(-1,1+a)-\zeta '(-1,1-a)
= -a\ln\left[{2\sin(\pi a)\over a}\right]+{1\over\pi}
\int_{0}^{\pi a}dz\,z\cot(z)\no$$
An alternative way of expressing this which is useful in the main text and
simplifies some of the expressions is
$$\zeta '(-1,a)-\zeta '(-1,1-a)
=-{1\over\pi}\int_{0}^{\pi a}\ln\left[2\sin(z)\right]\no$$
Let us now examine the consequences of decomposing over conjugacy
classes. Proceeding as for $\zeta '(0)$,
on decomposing over conjugacy classes  we  obtain the identity
$$\zeta '(-1)={1\over 12}\ln p+p\sum_{j=0}^{p-1}\zeta '(-1,1-{j\over
p})$$
For $p=2r+1$ this yields the identity
$$\eqalign{\zeta '(-1)&={1\over p^2-1}\left[p\sum_{l=1}^{(p-1)/2}\left[
{2l\over p}\ln\left[{2l\over p}\right]+{\left({2l\over
p}\right)}^2\right.\right.\cr
&\qquad\left.\left.
+\int_{0}^{2l/p}dy\,\left[\psi(1+y)+\psi(1-y)\right]\right]-{1\over
12}\ln p\right]\cr
}$$
i.e.
$$\eqalign{\zeta '(-1)&={1\over 6}-{1\over 4}\ln p\cr
&\qquad+{1\over p^2-1}\left[\sum_{l=1}^{(p-1)/2}\left[
2l\ln[2l]+p\int_{0}^{2l/p}dy\,\left[\psi(1+y)+\psi(1-y)\right]\right]
-{1\over 12}\ln p\right]\cr
}$$
This can be used numerically to verify that $\zeta '(-1)=0.16791$.
Now noting that
$$\int_{0}^{\pi(1-a)}dz\,\ln\left[2\sin(z)\right]=-\int_{0}^{\pi a}dz\,
\ln\left[2\sin(z)\right]$$
We find
$$\sum_{j=1}^{p-1}\int_{0}^{j\pi/p}dz\,\ln\left[2\sin(z)\right]=0$$
For $p$ even we find that by observing the two decompositions
$$\zeta '(-1,1+\alpha)=-\zeta '(-1,1+\alpha)\ln p
+p\sum_{l=0}^{p-1}\zeta '(-1,1-{l-\alpha\over p})$$
for $\alpha$ not necessarily integer and
$$\zeta '(-1,1+\alpha)=-\zeta(-1,1+\alpha)\ln p
+p\sum_{l=1}^{p}\zeta '(-1,{l+\alpha\over p})$$
valid for arbitrary $p$ can be used to simplify the expression for
$$\sum_{l=1}^{(p-2)/2}\int_{0}^{(2l+1)\pi/p}dz\,\ln\left[2\sin(z)\right]
={2\ln2\over p}-{1\over 2}\ln2-{4\over p}\ln[p/2]$$
These are used in simplifying the expression for the torsion in the case of $p$
even.
\subsec{ Relations involving $\zeta '(-2,1+a)$ }
We now turn to $\zeta '(-2,1+a)$ which we obtain from
$$\zeta(s-2,1+a)=\sum_{n=1}^{\infty}{1\over {(n+a)}^{s-2}}\no$$
We again extract the divergent parts from the sum to obtain
$$\eqalign{\zeta(s-2,1+a)&=\zeta(s-2)-(s-2)a\zeta(s-1)
+{(s-1)(s-2)a^2\over 2}\zeta(s)\cr
&\qquad-{s(s-1)(s-2)\zeta(s+1)a^3\over 6}+\hat\zeta(s-2,1+a)\cr}\no$$
where
$$\eqalign{\hat\zeta(s-2,1+a)&=\sum_{n=1}^{\infty}\left[
{1\over{(n+a)}^{s-2}}-{1\over n^{s-2}}+{(s-2)a\over{n}^{s-1}}\right.\cr
&\qquad\left.-{(s-1)(s-2)a^2\over 2{n}^{s}} +{s(s-1)(s-2)a^3\over6
n^{s+1}}\right]\cr}\no$$
Thus
$$\zeta(-2,1+a)=-{a\over 6}-{a^2\over 2}-{a^3\over 3}\no$$
and differentiating with respect to $s$ and
evaluating at $s=0$ we find
$$\eqalign{\zeta '(-2,1+a)&=\zeta '(-2)
+\left[2\zeta '(-1)+{1\over12}\right]a
+\left[{3\over 4}-{1\over2}\ln[2\pi]\right]a^2 \cr
&-\left[{\gamma\over 3}-{1\over2}\right]a^3+\hat\zeta '(-1,1+a)\cr
}\no$$
We evaluate
$\hat\zeta '(-2,1+a)$ by differentiating with respect to $s$ first
and then performing the sum over $n$ obtaining
$$\eqalign{\hat\zeta '(-2,1+a)
&=-\sum_{n=1}^{\infty}\left[ {(n+a)}^2\ln[1+{a\over n}]
-n\,a-{3a^2\over2}-{a^3\over 3n}\right]\cr
&=\sum_{k=2}^{\infty}{{(-1)}^{k}a^{k+2}\over k+2}\zeta(k)
-2a\sum_{k=2}^{\infty}{{(-1)}^{k}a^{k+1}\over k+1}\zeta(k)
+a^2\sum_{k=2}^{\infty}{{(-1)}^ka^k\over k}\zeta(k)\cr }\no$$
Thus
using the summation formula \docref{sumnf} we have
$$\hat\zeta '(-2,1+a)
=\int_{0}^{a}{(y-a)}^2\left[\psi(1+y)-\psi(1)\right] \no$$
We therefore obtain
$$\eqalign{\zeta '(-2,1+a) &=\zeta '(-2)
+{1\over12}\left[24\zeta '(-1)+1\right]a
+{1\over4}\left[3-2\ln[2\pi]\right]a^2\cr
&\qquad+{1\over2}a^3 +\int_{0}^{a}dy\,{(y-a)}^2\psi(1+y)\cr }
\no$$

We note that our expression for $\zeta '(-2,1+a)$
implies
$$\eqalign{\zeta '(-2,1+a)+\zeta '(-2,1-a)
&=2\zeta '(-2)
+{1\over2}\left[3-2\ln[2\pi]\right]a^2\cr
&\qquad+\int_{0}^{a}dy {(y-a)}^2\left[\psi(1+y)-\psi(1-y)\right]\cr
}\no$$
Differentiating the functional relation
$$\xi(s)=\xi(1-s),\quad\hbox{ where }\xi(s)=\Gamma(s/2)\pi^{-s/2}\zeta(s)\no$$
with respect to $s$ we find for $s=-2$
\eqlabel{\zmntwo}
$$\zeta '(-2)=-{1\over 4\pi^2}\zeta(3)\no$$
Finally using \docref{psirln}  and \docref{zmntwo} we have
$$\eqalign{\zeta '(-2,1+a)+\zeta '(-2,1-a)
&=-{1\over 2\pi^2}\zeta(3)-a^2\ln\left[{2\sin(\pi a)\over
a}\right]\cr
&\qquad-{1\over\pi^2}\int_{0}^{\pi a}dz\,z (z-2\pi a)\cot(z)\cr
}\no$$

We can obtain an identity by decomposing over conjugacy classes
mod $p$, by noting that
$$\zeta(s-2)=p^{2-s}\sum_{j=0}^{p-1}\zeta(s-2,1-{j\over p})\no$$
implies
$$
\zeta '(-2)=p^2\sum_{j=0}^{p-1}\zeta(-2,1-{j\over p})\ln p
+p^2\sum_{j=0}^{p-1}\zeta '(-2,1-{j\over p})
\no$$
The first sum is a decomposition of $\zeta(-2)$ and therefore
zero.
If we decompose $j$ into odd and even elements, by setting $j=2l+1$ and
$j=2l$ respectively, we find for $p$ odd i.e. $p=2r+1$
$$\zeta '(-2)=p^2\zeta '(-2)+p^2\sum_{l=1}^{r}
\left[\zeta '(-2,1-{2l\over p})+\zeta '(-2,1-{2l+1\over p})\right]
\no$$
which can be rewritten as either
$$\zeta '(-2)=p^2\zeta '(-2)
+p^2\sum_{l=1}^{r}\left[\zeta '(-2,1-{2l\over p})
+\zeta '(-2,{2l\over p})\right]
\no$$
or
$$\zeta '(-2)=p^2\zeta '(-2)+p^2\sum_{l=0}^{r-1}
\left[\zeta '(-2,{2l+1\over p})+\zeta '(-2,1-{2l+1\over p})\right]
\no$$
Therefore we have that
$$\zeta '(-2)=p^3\zeta
'(-2)-\sum_{l=1}^{r}\left[4l^2\ln\left[2\sin({2l\pi\over p})\right]
+{p^2\over\pi^2}\int_{0}^{2l\pi/p}dz\,
z(z-{4l\pi\over p})\cot(z)\right]\no$$
which on using \docref{zmntwo} gives
$${(1-p^3)\over 4\pi^2}\zeta(3)=
\sum_{l=1}^{r}\left[4l^2\ln\left[2\sin({2l\pi\over p})\right]
+{p^2\over\pi^2}\int_{0}^{2l\pi/p}dz\,
z(z-{4l\pi\over p})\cot(z)\right]\no$$
or
$$\eqalign{\zeta '(-2)&=p^3\zeta '(-2)-\sum_{l=0}^{r-1}\left[
{(2l+1)}^2\ln\left[2\sin({(2l+1)\pi\over p})\right]\right.\cr
&\qquad\left.+{p^2\over \pi^2}\int_{0}^{(2l+1)\pi/p}dz\,
z(z-{2(2l+1)\pi\over p})\cot(z)\right]\cr}
\no$$
which gives
$$\eqalign{{(1-p^3)\over4\pi^2}\zeta(3)&=\sum_{l=0}^{r-1}
\left[{(2l+1)}^2\ln\left[2\sin({(2l+1)\pi\over p})\right]\right.\cr
&\qquad\left.+{p^2\over \pi^2}\int_{0}^{(2l+1)\pi/p}dz\,
z(z-{2(2l+1)\pi\over p})\cot(z)\right]\cr}
\no$$
We can equally establish
that
$${(1-p^3)\over 4\pi^2}\zeta(3)=
\sum_{j=1}^{r}\left[j^2\ln\left[2\sin({j\pi\over p})\right]
+{p^2\over\pi^2}\int_{0}^{j\pi/p}dz\,
z(z-{2j\pi\over p})\cot(z)\right]\no$$

Finally we tabulate here for future reference some useful identities
regarding the Hurwitz and Riemann zeta functions
$$\zeta(s,a)={a}^{-s}+\zeta(s,1+a)\no$$
$$\eqalign{
\zeta(-2)&=0\qquad\zeta(-1,1+a)-\zeta(-1,1-a)=-a\cr
&\zeta(0,a)+\zeta(0,1-a)=0\qquad\zeta(0,1+a)+\zeta(0,1-a)=-1\cr
\zeta '(0,1+a)&=\zeta '(0,a)+\ln a\qquad
\zeta '(-1,1+a)=\zeta '(-1,a)+a\ln a\cr
\zeta '(0,1+a)+\zeta '(0,1-a)
&=-\ln\left[{2\sin(\pi a)\over a}\right]\cr}\no$$
$$\eqalign{
\zeta '(-1,1+a)-\zeta '(-1,1-a)
&=-a\ln\left[{2\sin(\pi a)\over a}\right]
+{1\over\pi}\int_{0}^{\pi a}dz\,z \cot(z)\cr
\zeta '(-1,a)-\zeta '(-1,1-a)
&=-a\ln(2\sin\pi a)+{1\over\pi}\int_{0}^{\pi a}dz\, z
\cot(z)\cr
}\no$$
and
$$\eqalign{\zeta '(-2,1+a)+\zeta '(-2,1-a)
&=-{1\over 2\pi^2}\zeta(3)-a^2\ln\left[{2\sin(\pi a)\over
a}\right]\cr
&\qquad-{1\over\pi^2}\int_{0}^{\pi a}dz\,z (z-2\pi a)\cot(z)\cr
}\no$$
$$\eqalign{\zeta '(-2,a)+\zeta '(-2,1-a)
&=-{1\over 2\pi^2}\zeta(3)-a^2\ln\left[2\sin(\pi a)\right]
-{1\over\pi^2}\int_{0}^{\pi a}dz\,z (z-2\pi a)\cot(z)\cr
}\no$$
\par\vskip\baselineskip
\centerline{\bf References}
\vskip0.5\baselineskip
{ \par \noindent \par \hangindent \parindent \indent
\hbox to 0pt {\hss \fam \bffam \tenbf 1.\kern .5em }
\ignorespaces
  Franz W., {\accent "7F U}ber die Torsion einer {\accent "7F U}berdeckung,
J. Reine Angew. Math.,
{\fam \bffam \tenbf 173}, 245--254, (1935).
\par \vskip -0.8\baselineskip \noindent }
{ \par \noindent \par \hangindent \parindent \indent
\hbox to 0pt {\hss \fam \bffam \tenbf 2.\kern .5em }
\ignorespaces
  Ray D. B., Reidemeister torsion and the Laplacian on lens spaces,
Adv. in Math.,
{\fam \bffam \tenbf 4}, 109--126, (1970).
\par \vskip -0.8\baselineskip \noindent }
{ \par \noindent \par \hangindent \parindent \indent
\hbox to 0pt {\hss \fam \bffam \tenbf 3.\kern .5em }
\ignorespaces
  Ray D. B. and Singer I. M.,
R-torsion and the Laplacian on Riemannian manifolds, Adv. in Math.,
{\fam \bffam \tenbf 7}, 145--201, (1971).
\par \vskip -0.8\baselineskip \noindent }
{ \par \noindent \par \hangindent \parindent \indent
\hbox to 0pt {\hss \fam \bffam \tenbf 4.\kern .5em }
\ignorespaces
  Ray D. B. and Singer I. M.,
Analytic Torsion for complex manifolds, Ann. Math.,
{\fam \bffam \tenbf 98}, 154--177, (1973).
\par \vskip -0.8\baselineskip \noindent }
{ \par \noindent \par \hangindent \parindent \indent
\hbox to 0pt {\hss \fam \bffam \tenbf 5.\kern .5em }
\ignorespaces
  Cheeger J., Analytic torsion and the heat equation, Ann. Math.,
{\fam \bffam \tenbf 109}, 259--322, (1979).
\par \vskip -0.8\baselineskip \noindent }
{ \par \noindent \par \hangindent \parindent \indent
\hbox to 0pt {\hss \fam \bffam \tenbf 6.\kern .5em }
\ignorespaces
  M{\accent "7F u}ller W.,
Analytic torsion and the R-torsion of Riemannian manifolds, Adv. Math.,
{\fam \bffam \tenbf 28}, 233--, (1978).
\par \vskip -0.8\baselineskip \noindent }
{ \par \noindent \par \hangindent \parindent \indent \hbox to 0pt
{\hss \fam \bffam \tenbf 7.\kern .5em }
\ignorespaces  Schwarz A. S.,
The partition function of degenerate quadratic functional and Ray-Singer
invariants, Lett. Math. Phys., {\fam \bffam \tenbf 2}, 247--252, (1978).
\par \vskip -0.8\baselineskip \noindent }
{ \par \noindent \par \hangindent \parindent \indent
\hbox to 0pt {\hss \fam \bffam \tenbf 8.\kern .5em }
\ignorespaces
  Witten E.,
{\fam \itfam \tenit Quantum field theory and the Jones polynomial},
I. A. M. P. Congress, Swansea, 1988, {edited by: Davies I., Simon B. and
Truman A.}, Institute of Physics, (1989).
 \par \vskip -0.8\baselineskip \noindent }
{ \par \noindent \par \hangindent \parindent \indent
\hbox to 0pt {\hss \fam \bffam \tenbf 9.\kern .5em }
\ignorespaces
  Nash C.,
{\fam \itfam \tenit Differential Topology and Quantum Field Theory},
Academic Press, (1991).
 \par \vskip -0.8\baselineskip \noindent }
{ \par \noindent \par \hangindent \parindent \indent
\hbox to 0pt {\hss \fam \bffam \tenbf 10.\kern .5em }
\ignorespaces
  Birmingham D., Blau M., Rakowski M. and Thompson G.,
Topological field theory, Phys. Rep.,
{\fam \bffam \tenbf 209}, 129--340, (1991).
\par \vskip -0.8\baselineskip \noindent }
{ \par \noindent \par \hangindent \parindent \indent
\hbox to 0pt {\hss \fam \bffam \tenbf 11.\kern .5em }
\ignorespaces
  Batalin I. A. and Vilkovisky G. A.,
Quantisation of Gauge Theories with linearly independent generators,
Phys. Rev.,
{\fam \bffam \tenbf D28}, 2567--, (1983).
\par \vskip -0.8\baselineskip \noindent }
{ \par \noindent \par \hangindent \parindent \indent
\hbox to 0pt {\hss \fam \bffam \tenbf 12.\kern .5em }
\ignorespaces
  Batalin I. A. and Vilkovisky G. A.,
Existence theorem for gauge algebras, Jour. Math. Phys.,
{\fam \bffam \tenbf 26}, 172--, (1985).
\par \vskip -0.8\baselineskip \noindent }
{ \par \noindent \par \hangindent \parindent \indent
\hbox to 0pt {\hss \fam \bffam \tenbf 13.\kern .5em }
\ignorespaces
  Rolfsen D.,
{\fam \itfam \tenit Knots and Links}, Publish or Perish, (1976).
 \par \vskip -0.8\baselineskip \noindent }
{ \par \noindent \par \hangindent \parindent \indent
\hbox to 0pt {\hss \fam \bffam \tenbf 14.\kern .5em }
\ignorespaces
  Bott R. and Tu L. W.,
{\fam \itfam \tenit Differential Forms in Algebraic Topology},
Springer-Verlag, New York, (1982).
 \par \vskip -0.8\baselineskip \noindent }
{ \par \noindent \par \hangindent \parindent
\indent \hbox to 0pt {\hss \fam \bffam \tenbf 15.\kern .5em }
\ignorespaces  Jeffrey L.,
Chern-Simons-Witten invariants of lens spaces and torus bundles,
and the semiclassical approximation,
Commun. Math. Phys.,
{\fam \bffam \tenbf 147}, 563--604, (1992).
\par \vskip -0.8\baselineskip \noindent }
{ \par \noindent \par \hangindent \parindent \indent
\hbox to 0pt {\hss \fam \bffam \tenbf 16.\kern .5em }
\ignorespaces
  Witten E.,
On quantum gauge theories in two dimensions, Commun. Math. Phys.,
{\fam \bffam \tenbf 141}, 153--209, (1991).
\par \vskip -0.8\baselineskip \noindent }
{ \par \noindent \par \hangindent \parindent \indent
\hbox to 0pt {\hss \fam \bffam \tenbf 17.\kern .5em }
\ignorespaces
  Poorten Alfred van der,
A proof that Euler missed.... Ap{\accent 19 e}ry's proof of the
irrationality of $\zeta (3)$. An informal report.,
The Mathematical Intelligencer,
{\fam \bffam \tenbf 1}, 195--203, (1979).
\par \vskip -0.8\baselineskip \noindent
}
{\par \noindent \par \hangindent \parindent \indent
\hbox to 0pt{\hss \fam \bffam \tenbf 18.\kern .5em }
\ignorespaces  Beukers, unpublished, private communication from W. Nahm.
}

\bye